\documentclass{aa}  
\usepackage{graphicx}
\usepackage{ulem}
\usepackage{txfonts}
\usepackage[switch]{lineno}
\usepackage{xcolor}
\usepackage{hyperref}
\hypersetup{
  colorlinks   = true,
  urlcolor     = blue,
  linkcolor    = blue,
  citecolor    = blue 
}
\usepackage{txfonts}
\usepackage{natbib}
\usepackage{newtxtext,newtxmath}
\usepackage[T1]{fontenc}
\usepackage{threeparttable}
\usepackage{comment}
\usepackage{tabularx}
\usepackage{lineno}
\usepackage{subcaption}
\usepackage{graphicx}
\usepackage{amsmath}
\usepackage{caption}

\DeclareCaptionLabelFormat{boldfig}{\textbf{Fig.}\ \textbf{#2.}}
\DeclareCaptionLabelFormat{boldtable}{\textbf{Table}\ \textbf{#2.}}
\newcommand{\vela}{Vela~X$-$1\xspace}
\newcommand{\maxi}{\textrm{MAXI}\xspace}
\newcommand{\soft}{2.0 $-$ 4.0 keV }
\newcommand{\hard}{12.0 $-$ 20.0 keV }
\newcommand{\overall}{2.0 $-$ 20.0 keV }

\begin{document}
\title{Unveiling the intensity-dependent wake structure of Vela X$-$1 using \textrm{MAXI}/GSC}
\author{Abhisek Tamang\inst{1,2}\thanks{E-mail: abhisek@rrimail.rri.res.in (RRI \& IISc.)}, Kinjal Roy\inst{1}, Hemanth Manikantan\inst{1}, Ajith Balu\inst{1} and Biswajit Paul\inst{1}}
\institute{Raman Research Institute, C. V. Raman Avenue, Sadashivanagar, Bangalore, Karnataka - 560080, India\and
Indian Institute of Science, C. V. Raman Avenue, Bangalore, Karnataka - 560012, India}

\date{Received DD August 2024; accepted DD MM YYYY}   

\abstract 
{Vela X$-$1 is one of the first few high-mass X-ray binary (HMXB) pulsars to be discovered. In HMXBs with pulsars such as Vela X$-$1, the companion’s stellar wind is significantly affected by ionisation due to X-rays from the compact object. An isotopic stellar wind model alone cannot explain the orbital variation in the absorption column density in Vela X$-$1. A model describing a stream-like photoionisation wake trailing the neutron star has been previously implemented to explain the observed orbital column density variation.}
{We investigated the variability of the circumbinary environment at different intensity levels of the Vela X$-$1 and used a model similar to the above-mentioned stream-like photoionisation wake to explain the asymmetric absorption column density present in the source.}
{The 2.0 - 20.0 keV \textrm{MAXI}/GSC spectrum was well modelled with a comptonised continuum emission absorbed by local and interstellar material. We used $\sim$13 years of \textrm{MAXI}/GSC data to constrain the variations in the absorption column density in Vela X$-$1 obtained from orbital-phase resolved and intensity-and-orbital-phase resolved spectral analysis.}
{The long-term light curve of Vela X$-$1 shows orbit-to-orbit intensity level variations without any apparent super-orbital periodicity. The orbital-phase resolved spectroscopy in multiple intensity levels reveals asymmetric variation in absorption column density changes across the intensity levels.}
{We confirm that the orbital variation in the absorption column density in Vela X$-$1 cannot be modelled with a smooth stellar wind alone using $\sim 13$ years of \maxi/GSC data. It requires an additional component, such as a photoionisation wake or an accretion wake. The wake structure is found to be present across different intensity levels of the source, and the geometry of the wake depends on the intensity level. The long-duration \maxi/GSC data allowed us to vary different wake parameters to obtain the best-fit stellar wind parameters for the time-averaged intensity. These best-fit parameters closely reproduce the observed orbital variations in the absorption column density for different intensity levels of the source.}

\authorrunning{A. Tamang et. al}
\maketitle

\section{Introduction}\label{sec: introduction}
High-mass X-ray binaries (HMXBs) are a sub-class of X-ray binaries in which a compact object accretes matter from a massive ($>$ 10 $M_\odot$) main-sequence stellar companion predominantly via the capture of stellar wind. 4U 0900-40 (henceforth~\vela ) was discovered with the Uhuru satellite in 1967~\citep{discovery_of_source}. It is the prototypical super-giant high-mass X-ray binary (sg-HMXB) pulsar. \vela comprises a neutron star with a mass of $M_{NS} \sim 1.77 M_\odot$~\citep{ns_mass} and a B0.5Ib optical companion that has a mass of $M_B \sim 23 M_\odot$ and a radius of $R_* \sim 30 R_\odot$~\citep{van_kerkwijk}. The neutron star has a spin period of $P_s \sim 283.5$ s~\citep{spin_period}. It is an eclipsing HMXB with an orbital period of $\sim$ 9 days~\citep{orbital_period}. The distance to the source has been calculated to be $1.99^{+0.13}_{-0.11}$ kpc~\citep{review_kretschmar}. The neutron star of~\vela orbits the companion in a slightly eccentric $\sim 0.1$ \citep{bildsten_1997} orbit with an apastron distance of $\sim 50 R_\odot$ (\citealt{bildsten_1997}) from the companion.

The companion star in the~\vela binary system is known to have a mass loss rate of $\sim 10^{-6} M_\odot$yr$^{-1}$~\citep{Nagase_1986} with a wind terminal velocity of $v_\infty \sim 1100$ km sec$^{-1}$~\citep{Watanabe_2006}. The neutron star accretes some of the stellar wind material, leading to X-ray emission. The binary environment populated by the stellar wind of the companion reprocesses part of the emitted X-rays via scattering and fluorescence emission. The signatures of the reprocessing region are encoded in the X-rays that reach us in the form of prominent Iron K$\alpha$ emission lines~\citep{iron_line_paper1, iron_line_paper2, Watanabe_2006, iron_line_paper4, iron_line_paper5}.
Some of the emitted X-rays are photo-electrically absorbed by the material crossing our line of sight~\citep{Vern_cs}. The variation in the absorption column density in the binary orbit tells us about the distribution of the absorbing medium. \vela exhibits variability in both absorption \citep{Nagase_1986, Haberl_1990, Watanabe_2006, malacaria_2016, review_kretschmar} and overall flux \citep{kreykenbohm_2008, furst_2010, review_kretschmar} on different timescales, from months \citep{Haberl_1990, kreykenbohm_2008} down to seconds (Fig. 1a of \citealt{inoue_1984}, Fig. 4 of \citealt{boerner_1987}, Fig. 5 in \citealt{kreykenbohm_2008}). \vela is also known to have off-states occurring outside of eclipses with minimal or undetectable X-ray emissions. These off-states have been attributed to diverse factors such as low-density regions or accretion processes hindered by magnetospheric effects~\citep{inoue_1984, vela_rxte_1999, sidoli_2015, review_kretschmar}. Several studies have revealed that the stellar wind of the companion in~\vela is not symmetric~\citep{blondin_1990, doroshenko_footprints, continuum_cyclotron_absorption_diez_2022, observing_the_onset_diez_2023}. Hence, the smooth-stellar wind model~\citep{smooth_wind_cak75} cannot explain the variation in the absorption column density with the orbital phase. Such an effect was also seen in some other HMXBs, such as IGR J17252-3616 \citep{Manousakis_2012} and 4U 1538-52 \citep{rodes_2015}. It has been suggested that this effect is due to the formation of a photoionisation wake, an accretion wake, or both because of the influence of a neutron star on the stellar wind. A photoionisation wake is formed when the X-rays emitted from the neutron star slow down the ejected wind by photoionising the medium, which reduces the scattering cross-section in the region; hence, a dense region within the str$\Ddot{\text{o}}$mgren sphere trailing the neutron star is formed \citep{blondin_1990, vela_x1_photoionization1994}. An accretion wake is formed as the gravity of the neutron star focuses the stellar wind onto its centre of gravity, thereby creating an extra-dense structure that trails the neutron star. Recently, when our work was under review, \cite{abalo_2024} reported results from their study of orbit-to-orbit variability in the flux of \vela~using more than 14 years of \maxi/GSC data. They used the hardness ratio to quantify the X-ray-absorbing material and concluded that a combination of an accretion wake, a photoionisation wake, and the clumpy nature of the stellar wind, along with the inherent variability of the source, contributes to these variations stochastically.

To explain the asymmetric variation in the absorption column density in~\vela,~\cite{doroshenko_footprints} proposed a toy model to include the contribution of the photoionisation wake. In this work, we have studied how the asymmetric variation changes across three different intensity levels -- low, mid, and high -- by performing intensity-and-orbital-phase resolved spectroscopy. For this analysis, we have used the long-term data acquired with \maxi/GSC in the last $\sim 13$ years. We have also estimated the wind parameters governing the stellar wind with the absorption column density obtained from the orbital-phase resolved spectroscopy with a similar wake model. 

The paper is organised as follows. In Sect.~\ref{sec: instrument_and_obs}, we present the instrument and data analysis techniques used. The timing analysis of the overall light curve ($\S$\ref{sec: overall_light_curve}) and the intensity-resolved light curves ($\S$\ref{sec: intensity_resolved_light_curves}) are presented in Sect.~\ref{sec: timing_analysis}. In Sect.~\ref{sec: spectral_analysis}, we perform time-averaged spectroscopy ($\S$\ref{sec: time_averaged_spectroscopy}), orbital-phase resolved spectroscopy ($\S$\ref{sec: orbital_phase_resolved_spectroscopy}), intensity-resolved spectroscopy ($\S$\ref{sec: intensity_resolved_spectroscopy}), and intensity-and-orbital-phase resolved spectroscopy ($\S$\ref{sec: intensity_and_orbital_phase_resolved_spectroscopy}). In Sect.~\ref{sec: wind_modelling}, we describe a toy model to explain the variation in the observed absorption column density. Finally, the implications of the results are discussed in Sect.~\ref{sec: discussion}.
 
\section{Instrument and observation}\label{sec: instrument_and_obs} 
Monitor of All-Sky X-ray Image (\textrm{MAXI}) is an X-ray monitoring mission operated from the International Space Station~\citep{maxi}. The \textrm{MAXI} has two X-ray instruments; namely, the Gas Slit Camera (GSC) \citep{maxi_gsc} and the Solid-state Slit Camera (SSC)~\citep{maxi_ssc}. The \maxi/GSC instrument consists of 12 position-sensitive gas proportional counters that operate in the energy range 2.0 - 30.0 keV, and \textrm{MAXI}/SSC has charge-coupled devices (CCDs) that operate in the energy range 0.5 - 12 keV. These two instruments cover the entire sky in every ISS orbital period ($\sim$ 90 minutes) with fields of view of 1.$^\circ$5 $\times$ 160$^\circ$ and 1.$^\circ$5 $\times$ 90$^\circ$ for GSC and SSC, respectively.

For the current work, we have utilised \maxi/GSC data for~\vela from MJD 55200 to MJD 59791. The spectra and light curves were obtained from the \maxi\ on-demand website\footnote{\url{http://maxi.riken.jp/mxondem/}}. The timing resolution of all the light curves was taken as 0.0625d ($\sim 90$ min $\sim$ one ISS orbit). For orbital-phase resolved and intensity-resolved spectral analysis, we downloaded the data using good time interval files to select the relevant phase and intensity level, respectively.

\section{Timing analysis}\label{sec: timing_analysis}
\subsection{Overall light curve}\label{sec: overall_light_curve}
\begin{table}
\centering
\caption{Orbital ephemeris of~\vela from~\citet{kreykenbohm_2008}.}
\begin{threeparttable}
\begin{tabularx}{0.45\textwidth}{*{2}{>{\centering\arraybackslash}X}}
\hline
\hline
Orbital parameter                     & Value                  \\ \hline
P$_{\text{orb}}$          & 8.964357 $\pm$ 0.000029 days \\
T$_{\text{Mid-Eclipse}}$\tnote{a} & 55134.63704 MJD             \\ 

$\Omega$ & $152.59 \pm 0.92^\circ$ \\
$i$ & $ > 74^\circ$ \\
$e$ & $0.0898 \pm 0.0012$ \\ 
\hline
\end{tabularx}
\begin{tablenotes}
    \item[a] It is taken as phase zero \citep{malacaria_2016}.
\end{tablenotes}
\end{threeparttable}
\label{tab: ephermeris_table}
\end{table}

\begin{figure*}
  \centering
  \includegraphics[width=\textwidth]{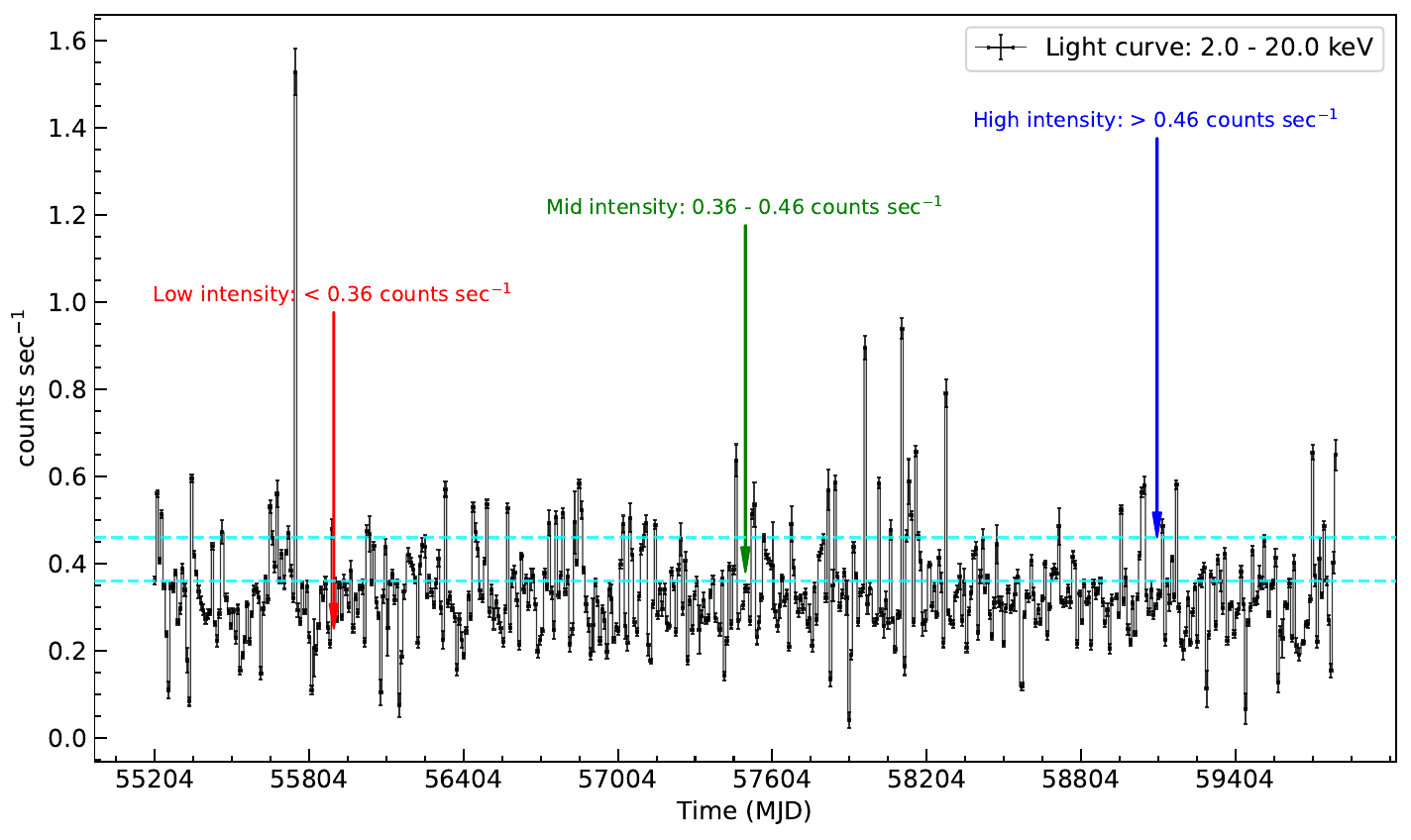}
  \caption{$\sim13$-year long-term \maxi/GSC light curve of~\vela with a bin size of one orbital period. The red, green, and blue arrows mark the low ($<0.36 $ counts sec$^{-1}$), mid (0.36 - 0.46 counts sec$^{-1}$), and high ($> 0.46$ counts sec$^{-1}$) intensity levels. The horizontal lines (cyan) separate the three intensity levels.}
  \label{fig:vela_x1_long_term_lc_2to20kev}
\end{figure*}

\begin{figure}
  \centering
  \includegraphics[width=0.5\textwidth]{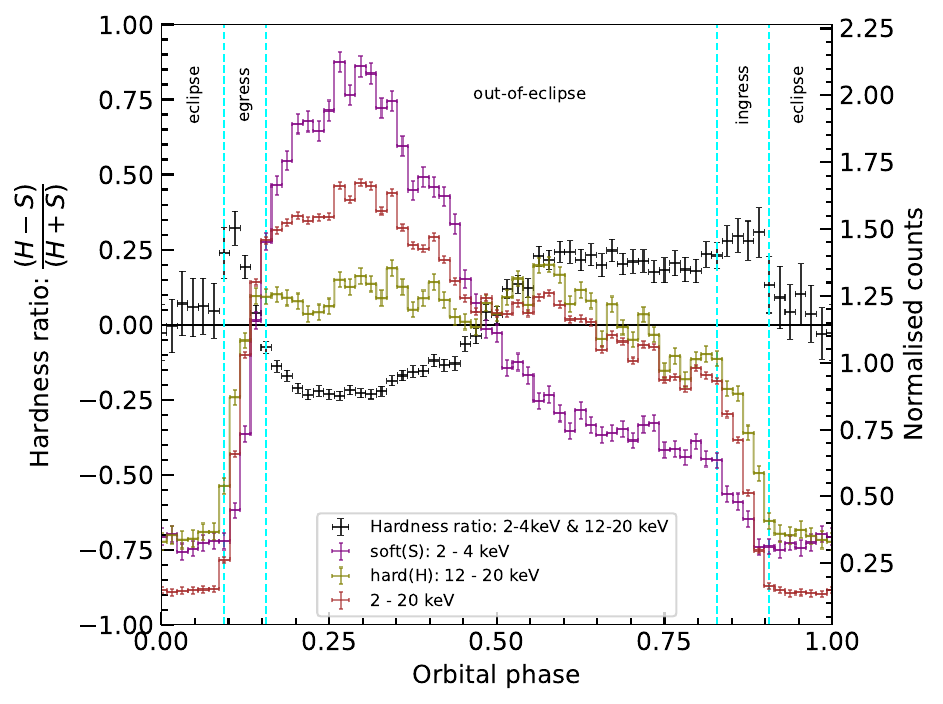}
  \caption{Energy-dependent orbital profiles 2.0 - 4.0 keV (purple), 12.0 - 20.0 keV (olive), and 2.0 - 20.0 keV (brown) are plotted with the bin size of 64 bins per interval. The hardness ratio (black) is plotted between 2.0 - 4.0 keV (soft) and 12.0 - 20.0 keV (hard) energy bands. The vertical lines (cyan) divide the orbital profiles into eclipse, egress, OOE, and ingress.}
  \label{fig:vela_x1_total_int_hardness}
\end{figure}

The long-term light curve of~\vela obtained in the \overall range is plotted in Fig.~\ref{fig:vela_x1_long_term_lc_2to20kev} with a bin size of one orbital period. The light curve shows long-term variations (Fig.~\ref{fig:vela_x1_long_term_lc_2to20kev}) with no apparent periodicity. The orbital period of~\vela was determined from the \maxi/GSC light curve using the epoch folding tool, \texttt{efsearch}\footnote{\url{https://heasarc.gsfc.nasa.gov/xanadu/xronos/examples/efsearch.html}} to be $8.96432 \pm 0.00029$ days, consistent with~\cite{kreykenbohm_2008}. The other orbital parameters used in this paper are given in Table~\ref{tab: ephermeris_table}. The orbital intensity profile in \overall energy band is plotted in brown in Fig.~\ref{fig:vela_x1_total_int_hardness}. The orbital profile of~\vela is asymmetric in nature, with a significantly higher count rate before an orbital phase of $\sim$ 0.4 \citep{doroshenko_footprints}. 

To study the variation in the orbital profile with energy, we downloaded the source light curves in soft and hard energy bands of \soft (soft) and \hard keV (hard), respectively. The soft and hard energy bands orbital profiles are plotted in Fig.~\ref{fig:vela_x1_total_int_hardness} with 64 bins per period, and the hardness ratio between the two bands is also shown. The count rate in the soft energy band decreases after an orbital phase of 0.4, whereas the hard X-ray profile is nearly flat. The hardness ratio also shows higher values during eclipse-ingress and eclipse-egress. There is an excess of hard photons after an orbital phase of 0.4. A similar variation in the hardness ratio was reported by \cite{continuum_cyclotron_absorption_diez_2022, observing_the_onset_diez_2023} with \textrm{NuSTAR} (2022) and \textrm{XMM}-Newton (2023) observations. A similar variation in the averaged hardness ratio between the energy bands 2.0 - 4.0 keV and 4.0 - 10.0 keV was shown by \cite{abalo_2024}, even though the hardness ratio varies from orbit-to-orbit.

\subsection{Intensity resolved light curves}\label{sec: intensity_resolved_light_curves}
\begin{figure}
  \centering
  \includegraphics[width=0.5\textwidth]{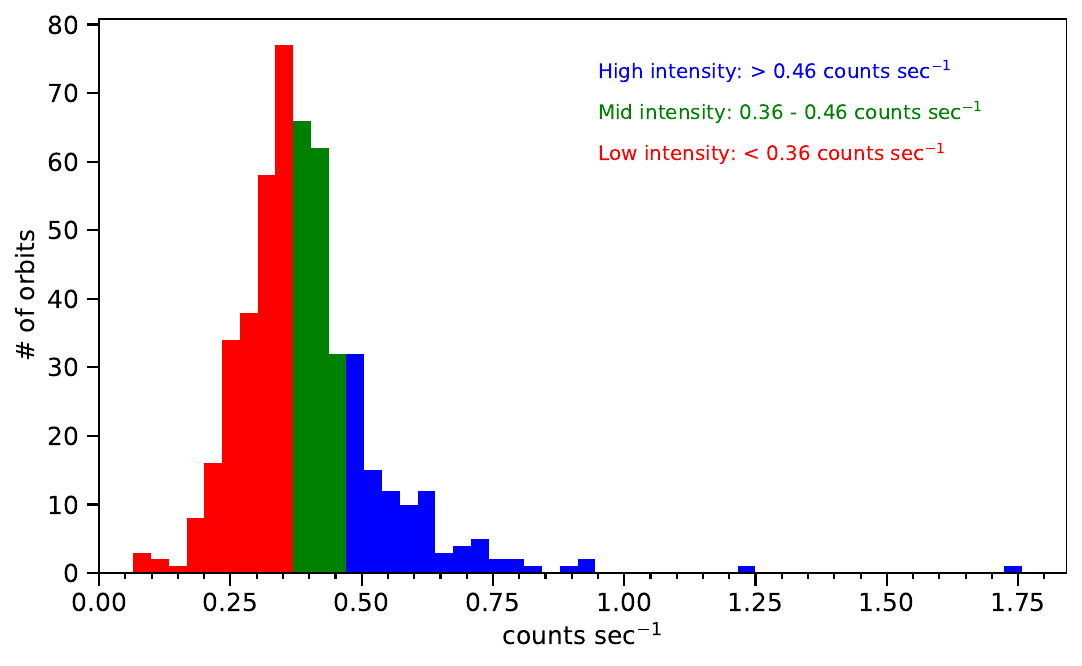}
  \caption{Histogram of total intensity per orbit grouped into three intensity levels. The X axis represents the \maxi/GSC counts rate per orbit. The Y axis represents the number of orbits. The counts rate range for the low (red), mid (green), and high (blue) intensity levels are $< 0.36$ counts sec$^{-1}$, $0.36 - 0.46$ counts sec$^{-1}$, and $> 0.46$ counts sec$^{-1}$, respectively. The total number of orbits in the three intensity levels are 222, 161, and 111, respectively.}
  \label{fig:vela_x1_orbits_histogram_3int_levels}
\end{figure}

\begin{figure}
  \centering
  \includegraphics[width=0.45\textwidth]{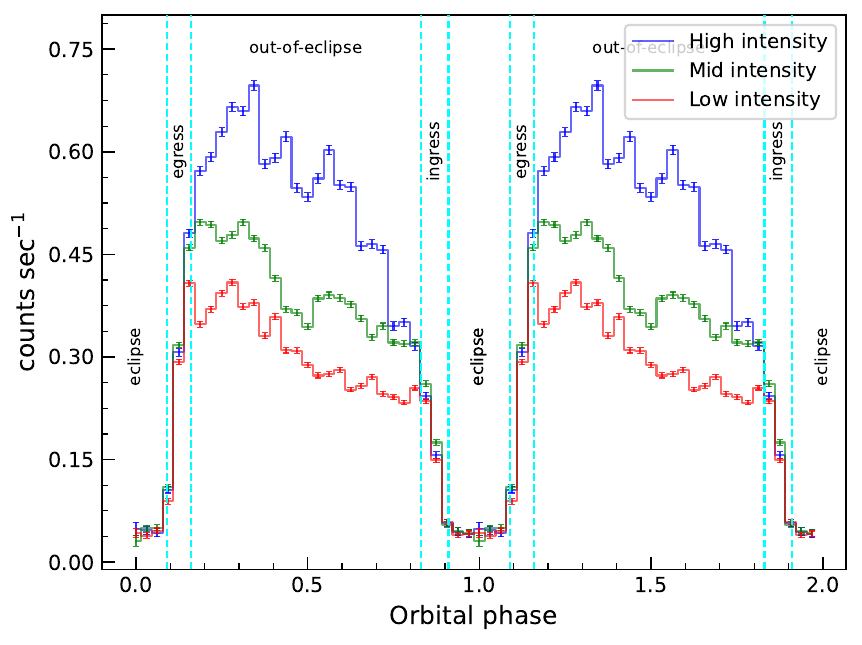}
  \caption{\overall orbital profiles of Vela X-1 are plotted for low (red), mid (green), and high (blue) intensity levels with the bin size of 32 bins per interval. The vertical lines (cyan) divide the orbital profiles into eclipse, egress, OOE, and ingress.}
  \label{fig:int_resolved_orbital_profiles}
\end{figure}

\begin{figure}
  \centering
  \includegraphics[width=0.48\textwidth]{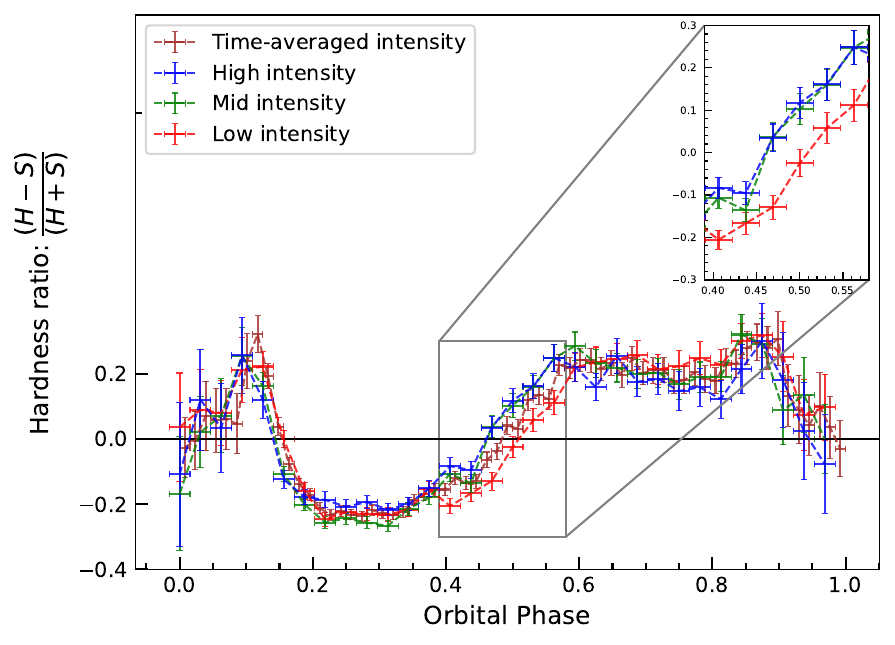}
  \caption{Hardness ratio plots between the soft (S; 2.0-4.0 keV) and hard (H; 12.0-20.0 keV) energy bands for the low (red), mid (green), and high (blue) intensity levels. The inset in the upper right corner shows that the hardness ratio in the low-intensity level rises slowly compared to the high-intensity level in the phase range 0.4 - 0.6.}
  \label{fig:vela_x1_int_resolved_hardness}
\end{figure}

The long-term light curve of~\vela in Fig.~\ref{fig:vela_x1_long_term_lc_2to20kev} suggests that the source intensity varies from orbit to orbit. The long-term \maxi/GSC data has covered approximately 512~\vela binary orbits during the observation period. We are interested in intensity variation in the out-of-eclipse (OOE) region; hence, we selected approximately the phase range $0.15 - 0.80$ (where $\phi = 0$ is the mid-eclipse) to study the orbit-to-orbit intensity variation. The average count rate per orbit for all orbits in the aforementioned phase range was calculated, and its histogram is plotted in Fig.~\ref{fig:vela_x1_orbits_histogram_3int_levels}. The distribution suggests that the average intensity indeed varies by a few factors from orbit to orbit; the variable nature of~\vela over different timescales has been reported in the literature~\citep{kreykenbohm_2008, furst_2010, malacaria_2016}. The orbits of~\vela were divided into three intensity levels. While dividing, we ensured that the sum of the averaged count rate of all the orbits in each intensity region was comparable. In the present case, the sum is around 65 counts. Under this condition, the following threshold values for count rates were obtained for the different intensity levels: low,  $< 0.36$ counts sec$^{-1}$ with 222 orbits; mid, $0.36 - 0.46$ counts sec$^{-1}$ with 161 orbits; and high, $> 0.46$ counts sec$^{-1}$ with 111 orbits, respectively. The orbital profiles in the energy band of \overall in the low, mid, and high-intensity levels are plotted in Fig.~\ref{fig:int_resolved_orbital_profiles} with 32 bins per period. The orbital intensity profile in the high-intensity level has some sub-structures. It is not known what causes the multiple peaks in the high-intensity level, but such peaks are also observed in the orbital intensity profile of Vela X-1 made in the hard X-ray band using the INTEGRAL/ISGRI (17 - 150 keV) light curve (Fig. 3 of \citealt{falanga_2015}). In the mid-state, there is a dip between phase 0.4 $-$ 0.5 on top of a decaying profile. Meanwhile, in the low state, there is a gradual decay of the count rate. The intensity resolved hardness ratio plot is given in Fig.~\ref{fig:vela_x1_int_resolved_hardness}; the hardness ratio between \soft and \hard in the low-intensity level rises slowly compared to the high-intensity level.

\section{Spectral analysis}\label{sec: spectral_analysis}
The variation in the hardness ratio throughout the orbit and between different intensity levels indicates changes in the spectral parameters. We performed time-averaged spectroscopy and orbital-phase resolved spectroscopy to probe the orbital variation in spectral parameters in the overall spectrum and in different intensity levels. The errors on best-fit spectral parameters in Table~\ref{tab: time_averaged_and_intensity_resolved_spectroscopy} are quoted at a 90\% confidence level. The spectral analysis was carried out using \texttt{XSPEC} v: 12.13.0c~\citep{XSPEC}. The elemental abundance was taken from~\citet{tbabs_wilms_2000} and the photoelectric cross-sections from~\citet{Vern_cs}. 

\subsection{Time-averaged spectroscopy}\label{sec: time_averaged_spectroscopy}
\begin{figure}
  \centering
  \includegraphics[width=0.47\textwidth]{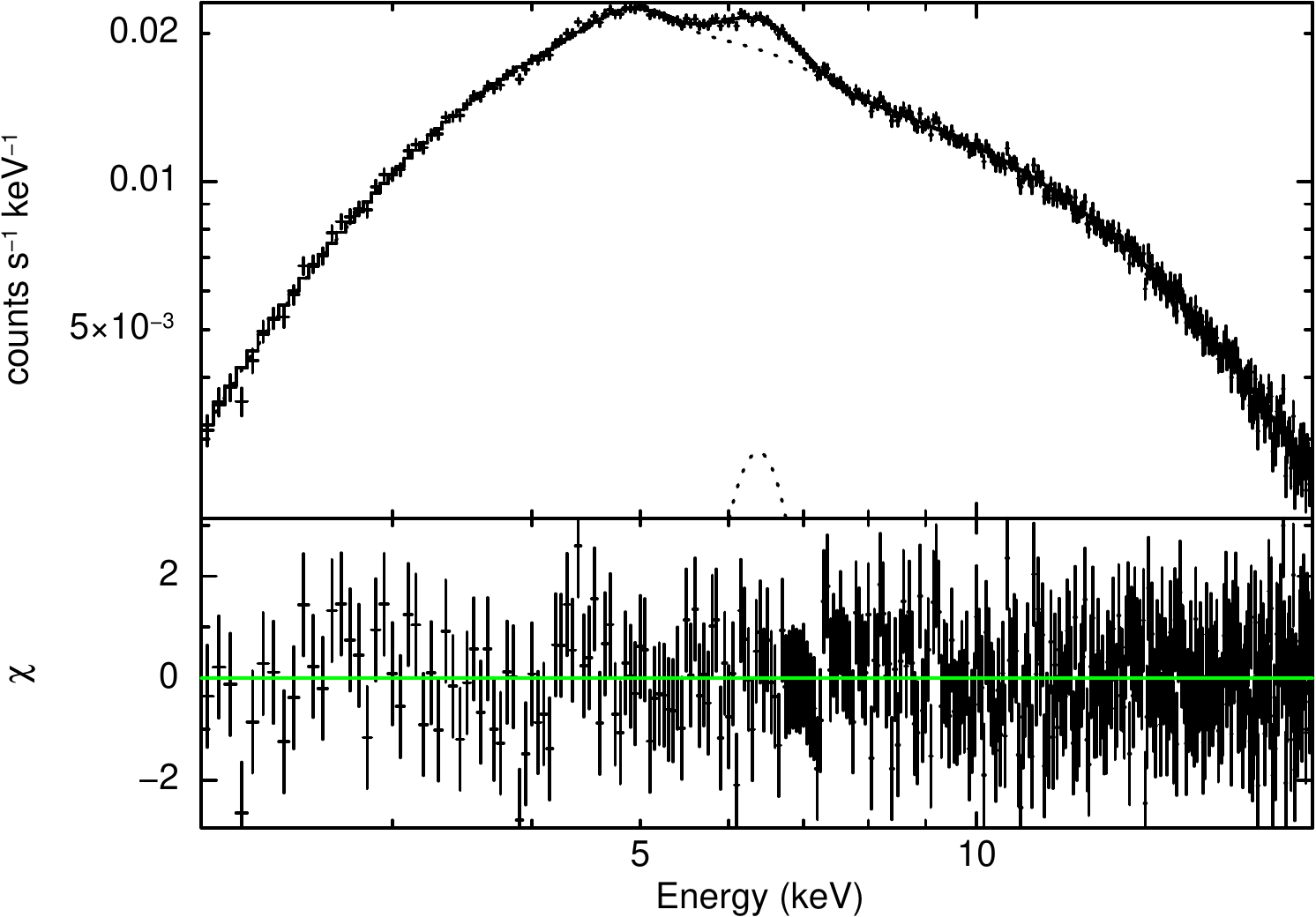}
  \caption{Time-averaged spectrum of~\vela in \overall energy band obtained from \maxi/GSC. Top panel: Data (black crosses) and best-fit model (dashed line) components. Bottom panel: Deviation of the data from the best-fitted model spectrum.} 
  \label{fig:vela_x1_time_averaged_spectrum}
\end{figure}

Different phenomenological models have been used to study Vela X-1. In particular, some have described its continuum emission by a cut-off power law (e.g. \citealt{malacaria_2016}) or a power-law with a Fermi Dirac cut-off (e.g. \citealt{continuum_cyclotron_absorption_diez_2022, observing_the_onset_diez_2023}). In this work, we use a physically motivated Comptonised continuum model, which allows us to compare our results with the ones reported earlier by \cite{doroshenko_footprints}. Hence, our final best-fit model is tbabs*pcfabs(CompST $+$ Gaussian), where \texttt{CompST} is the inverse-comptonisation of cool photons by the hot electrons \citep{compst_model}, \texttt{tbabs} \citep{tbabs_wilms_2000} models the photoelectric absorption by the Galactic ISM along the line of sight, \texttt{pcfabs} was used for incorporating the contribution from the variable local absorption by the stellar wind, and there is a prominent Fe K$\alpha$ line at $\sim$6.4 keV that was modelled with a single Gaussian emission line \texttt{(Gaussian)}. The Galactic absorption component modelled by \texttt{tbabs} was frozen to the Galactic value ($\sim 0.37 \times 10^{22}$ cm$^{-2}$) along the line of sight \citep{nH_calculator_heasarc}. The best-fit parameter values along with error estimates at a 90\% confidence level are given in Table~\ref{tab: time_averaged_and_intensity_resolved_spectroscopy} and are consistent with results reported by \cite{doroshenko_footprints} within the error bars. The corresponding spectrum, best-fit model, and residuals for the best-fit model are shown in Fig.~\ref{fig:vela_x1_time_averaged_spectrum}.

\begin{table*}
\renewcommand{\arraystretch}{1.5}
\centering
\caption{Best-fit parameters obtained from the time-averaged intensity and intensity-resolved spectroscopy.}
\begin{threeparttable}
\begin{tabularx}{1.0\textwidth}{XXXXXXX}
\hline
\hline
                                   &                                                &                                      &                         & Best Fit Values            &                            &                   \\ \hline
Component                          & Parameter                                      & Units                                & Time-averaged intensity       & Low intensity              & Mid intensity              & High intensity             \\ \cline{4-7} 
\texttt{tBabs}                     & $N_\text{H$_\text{gal}$}$                      & 10$^{22}$ cm$^{-2}$                  & 0.37\tnote{a}           & 0.37\tnote{a}              & 0.37\tnote{a}              & 0.37\tnote{a}              \\
\texttt{pcfabs}                    & $N_\text{H$_\text{pcfabs}$}$ & 10$^{22}$ cm$^{-2}$ & 17.02$\pm 0.89$          & 16.7$\pm 1.4$ & 16.0$\pm 1.3$    & 18.4$_{-1.4}^{+1.5}$                                 \\
                                   & CvrFract                                       &                                      & 0.83$\pm 0.01$  & 0.83$\pm 0.01$     & 0.82$\pm 0.01$     & 0.83$\pm 0.01$     \\
\texttt{CompST}                    & kT                                             & keV                                  & 5.96$_{-0.18}^{+0.20}$  & 5.69$_{-0.28}^{+0.34}$     & 6.07$_{-0.28}^{+0.33}$     & 6.08$_{-0.28}^{+0.34}$     \\
                                   & $\tau$                                         &                                      & 13.02$\pm 0.38$ & 12.83$\pm 0.62$    & 13.08$_{-0.56}^{+0.55}$    & 13.31$\pm 0.62$    \\
                                   & Norm                                           &                                      & 0.27$\pm 0.01$  & 0.24$\pm 0.02$     & 0.28$\pm 0.02$     & 0.35$\pm 0.03$     \\
\texttt{Gaussian}                  & $E_\text{line}$                                & keV                                  & 6.35$\pm 0.04$  & 6.35\tnote{b}              & 6.35\tnote{b}              & 6.35\tnote{b}              \\
                                   & $\sigma_\text{line}$                           & keV                                  & $< 0.26$  & 0.15\tnote{b}              & 0.15\tnote{b}              & 0.15\tnote{b}              \\
                                   & Norm                                           & $10^{-3}$ photons cm$^{-2}$ sec$^{-1}$ & 4.50$_{-0.43}^{+0.49}$  & 3.62$_{-0.46}^{+0.45}$     & 4.51$_{-0.55}^{+0.54}$     & 6.84$_{-0.80}^{+0.78}$     \\
                                   & Eq. width                                      & eV                                   & 182$_{-20}^{+21}$    & 178$_{-23}^{+27}$ & 174$\pm 24$ & 201$_{-25}^{+26}$ \\
Flux$_\text{2-20 keV}$             &                                                & 10$^{-9}$ ergs cm$^{-2}$ sec$^{-1}$    & 3.50$_{-0.02}^{+0.01}$  & 2.76$_{-0.04}^{+0.01}$     & 3.76$_{-0.04}^{+0.02}$     & 4.90$_{-0.06}^{+0.02}$     \\ \hline
$\chi^2 /$dof                      &                                                &                                      & 320.84$/$351            & 342.76/353                 & 352.61/353                 & 359.03/353                 \\ 
\hline
\hline
\end{tabularx}
\begin{tablenotes}
    \item[a] Frozen to the Galactic value along the line of sight.
    \item[b] These values were frozen to their time-averaged values.
\end{tablenotes}
\end{threeparttable}
\label{tab: time_averaged_and_intensity_resolved_spectroscopy}
\end{table*}

\subsection{Orbital-phase resolved spectroscopy}\label{sec: orbital_phase_resolved_spectroscopy}
\begin{figure}
  \centering
  \includegraphics[width=0.5\textwidth]{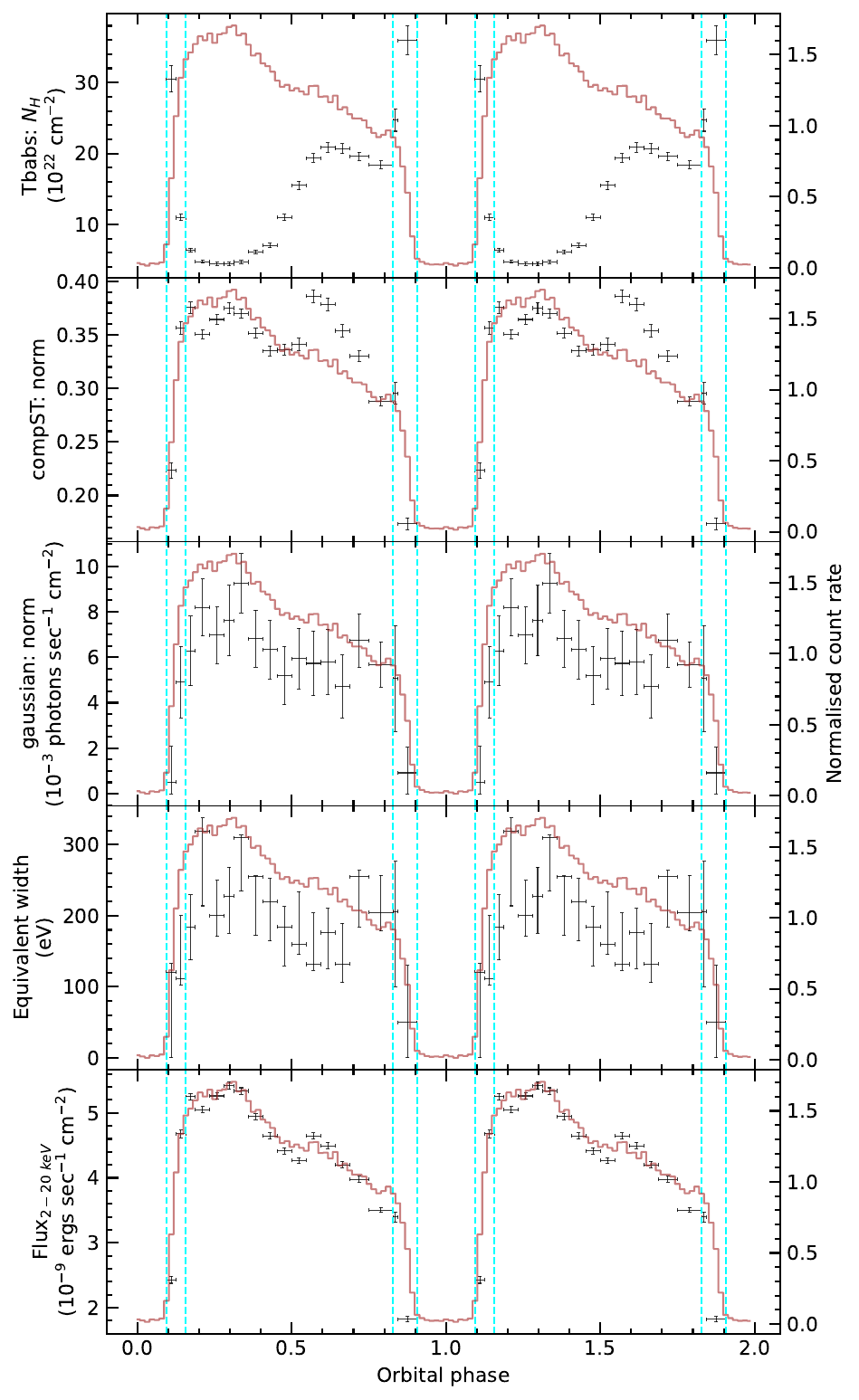}
  \caption{Variation in spectral parameters (black) obtained from orbital-phase resolved spectroscopy. The error bars are at the 90\% confidence level. The orbital profile (brown) is plotted in the 2.0 - 20.0 keV energy band. The vertical lines (cyan) are drawn to denote the four different regions (eclipse, egress, OOE, and ingress).}
  \label{fig: vela_x1_orbital_phase_resolved_spectroscopy}
\end{figure}

In order to study the orbital variation in the hardness ratio (Fig.~\ref{fig:vela_x1_total_int_hardness}, Fig.~\ref{fig:vela_x1_int_resolved_hardness}) further, we performed orbital-phase resolved spectroscopy. We divided the \vela orbital profile (Fig.~\ref{fig:vela_x1_total_int_hardness}) into three regions, namely the eclipse-egress (egress hereafter), OOE, and eclipse-ingress (ingress hereafter), with corresponding phase ranges of 0.09 - 0.16, 0.16 - 0.83, and 0.83 - 0.91, respectively, after visual inspection of the hardness ratios and orbital profiles. The three regions were further divided into multiple orbital phase bins, with the OOE into 14 orbital phase bins, and the egress and ingress into two orbital phase bins each, ensuring that each phase bin contains approximately the same total number of source photons. Using orbital ephemeris (given in Table~\ref{tab: ephermeris_table}), we performed joint fitting of all 18 spectra with the model \texttt{tbabs*(CompST + Gaussian)}. Here, we removed \texttt{pcfabs}, considering the reduced total counts for orbital-phase resolved spectroscopy, and thus avoided over-fitting the spectra, similar to what was done in \cite{doroshenko_footprints}. We let \texttt{tbabs} alone account for both local and Galactic absorption. We fixed the temperature ($T$) and optical depth ($\tau$) of \texttt{CompST} and the line centre and width of the iron emission line to their respective time-averaged values (given in Table~\ref{tab: time_averaged_and_intensity_resolved_spectroscopy}), as they could not be constrained in individual orbit phases due to limited photon statistics. The variation in different spectral parameters with the orbital phase is shown in Fig.~\ref{fig: vela_x1_orbital_phase_resolved_spectroscopy}. The top panel in Fig.~\ref{fig: vela_x1_orbital_phase_resolved_spectroscopy} shows the variation in the absorption column density with the orbital phase. The absorption column density starts increasing after phase 0.4 as in the case of the hardness ratio. Such a variation in~\vela was previously reported in the literature~\citep{doroshenko_footprints, continuum_cyclotron_absorption_diez_2022, observing_the_onset_diez_2023}.

\subsection{Intensity resolved spectroscopy}\label{sec: intensity_resolved_spectroscopy}
\begin{figure}
  \centering
  \includegraphics[width=0.45\textwidth]{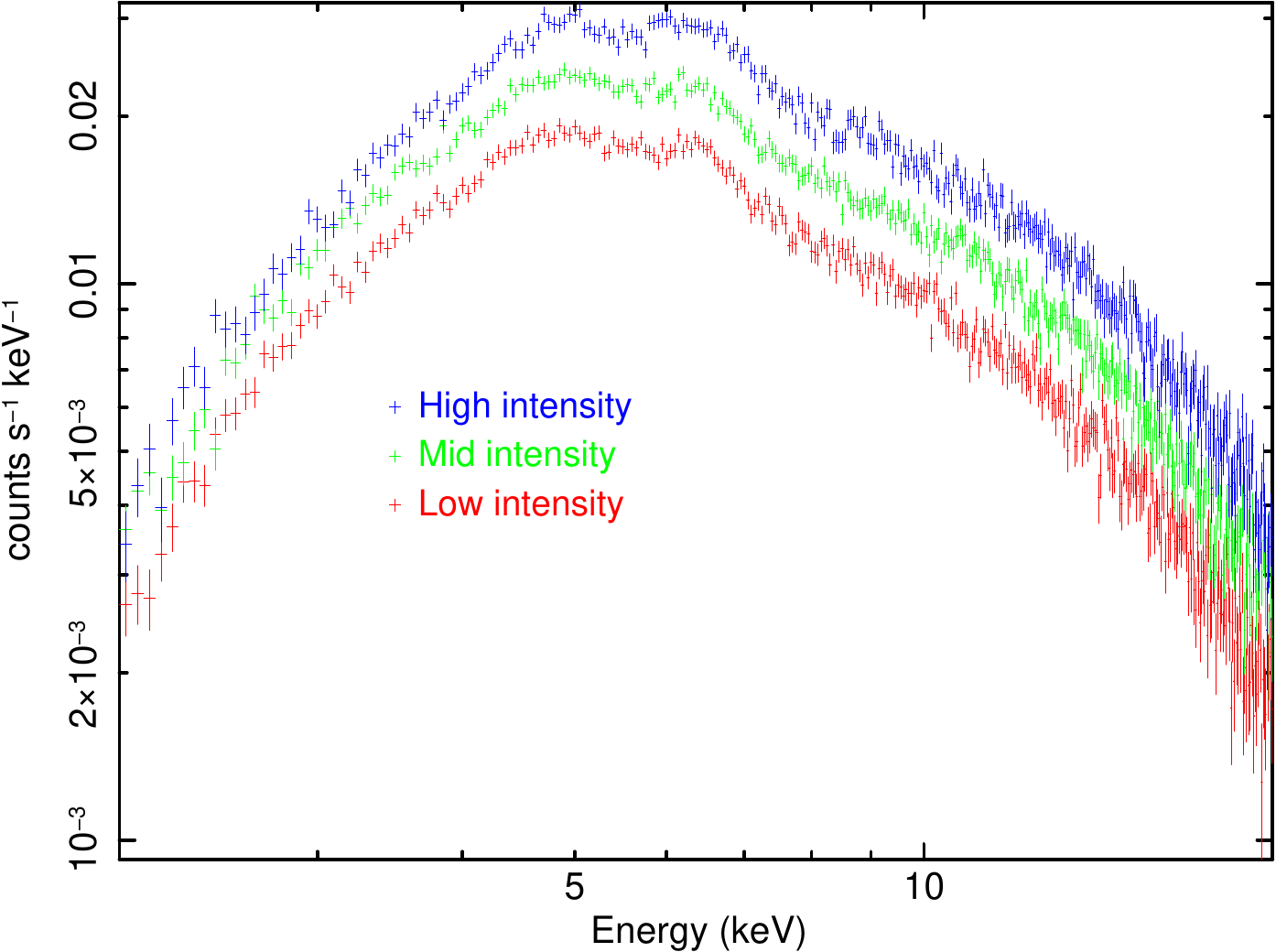}
  \caption{Three intensity-resolved spectra in the \overall energy band.}
  \label{fig: vela_x1_int_resolved_spectra}
\end{figure}

The asymmetric distribution observed in the absorption column density depicted in Fig.~\ref{fig: vela_x1_orbital_phase_resolved_spectroscopy} and varying rise of the hardness ratio across different intensity levels in Fig.~\ref{fig:vela_x1_int_resolved_hardness} prompts us to investigate similar variations across different intensity levels. Additionally, the fluctuating count rates from orbit to orbit illustrated in Fig.~\ref{fig:int_resolved_orbital_profiles} suggest potential differences in spectral parameters across different intensity levels.

We did intensity-resolved spectroscopy by fitting each spectrum to the same spectral model, \texttt{tbabs*pcfabs*(CompST $+$ Gaussian)}, as the time-averaged spectrum (Sect.~\ref{sec: time_averaged_spectroscopy}). Similar to time-averaged spectroscopy, we fixed \texttt{tbabs} $N_H$ to the Galactic absorption column density along the line of sight, while allowing the local absorption parameters (\texttt{pcfabs}) to vary. Due to constraints imposed by reduced photon counts, the emission line centre and width were fixed to their respective values determined from the time-averaged spectrum. The rest of the parameters were kept free. The best-fit parameters' values obtained from the fits for three different intensity levels are given in Table~\ref{tab: time_averaged_and_intensity_resolved_spectroscopy} and the corresponding spectra are shown in Fig.~\ref{fig: vela_x1_int_resolved_spectra}. From Fig.~\ref{fig: vela_x1_int_resolved_spectra}, the three intensity states do not show significant changes in spectral shape, but the total emission strength has increased from the low- to the high-intensity level. The iron line is present in all three spectra. In Table~\ref{tab: time_averaged_and_intensity_resolved_spectroscopy}, along with an increase in the intensity levels, normalisations of the continuum (\texttt{CompST}) and the iron emission line (\texttt{Gaussian}) are found to increase.

\subsection{Intensity-and-orbital-phase resolved spectroscopy}\label{sec: intensity_and_orbital_phase_resolved_spectroscopy}
\begin{figure}
  \centering
  \includegraphics[width=0.5\textwidth]{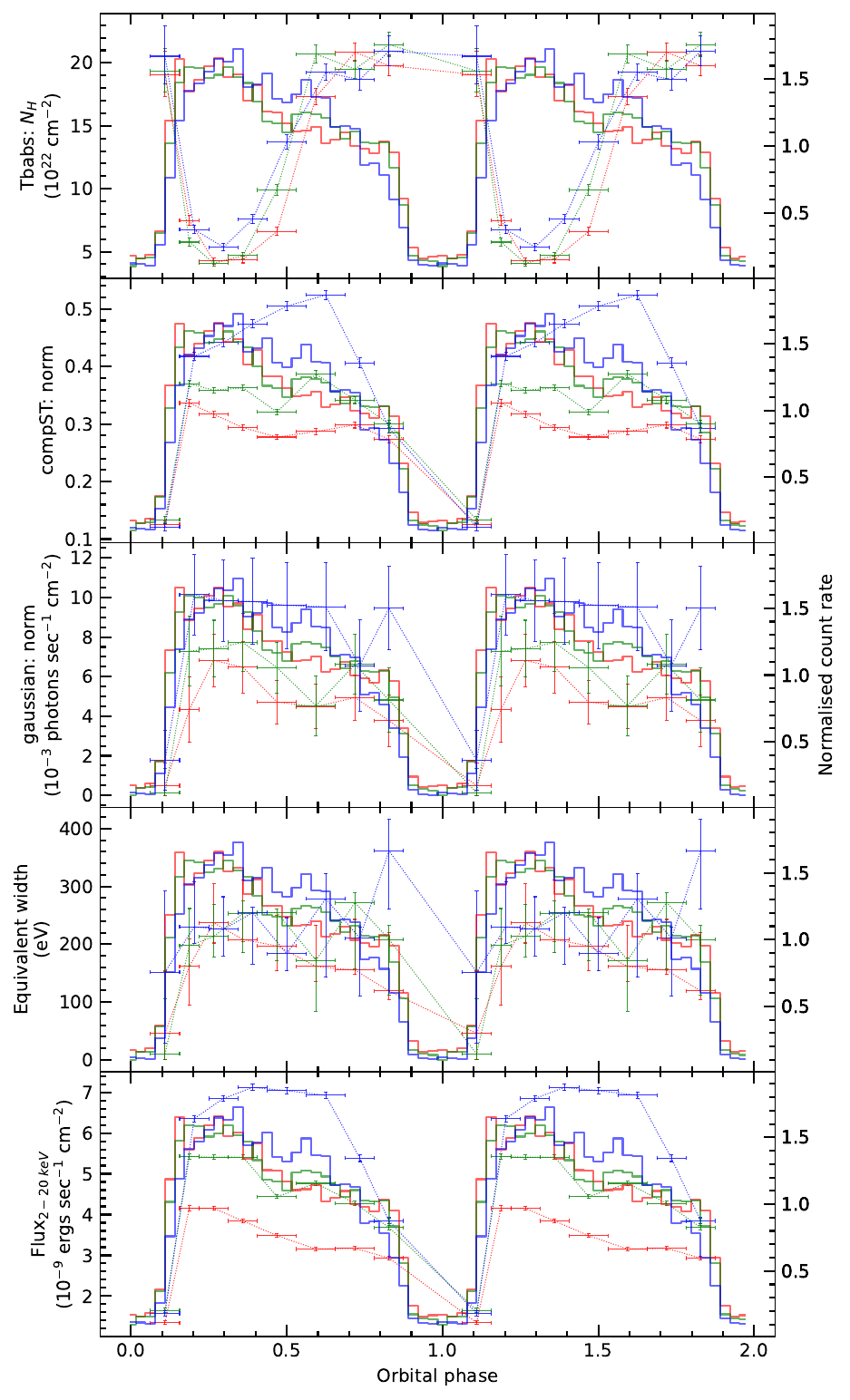}
  \caption{Variation in spectral parameters obtained from intensity-and-orbital-phase resolved spectroscopy for the low (red), mid (green), and high (blue) intensity levels. The error bars are at a 90\% confidence level. The orbital profiles for low (red), mid (green), and high (blue) intensity levels are plotted in the \overall band.}
  \label{fig: vela_x1_int_phase_resolved_spectroscopy}
\end{figure}

We also performed orbital-phase resolved spectroscopy on each intensity level to probe the variation in patterns in the spectral parameters as a function of the orbital phase. The net count rate was reduced in each intensity level; hence, the orbital OOE region was subdivided into six spectral bins, along with one bin each for egress and ingress. We ensured an approximately equal number of counts for all phase bins. We conducted joint fitting of eight spectra of the three intensity levels using the same spectral model utilised for orbital-phase resolved spectroscopy described in Sect.~\ref{sec: orbital_phase_resolved_spectroscopy}. The temperature ($T$) and optical depth ($\tau$) of \texttt{CompST}, as well as the centre and width of iron emission lines, were fixed to their respective intensity resolved values. The results, depicted in Fig.~\ref{fig: vela_x1_int_phase_resolved_spectroscopy}, illustrate how spectral parameters vary with the orbital phase across the three intensity levels. Notably, the asymmetric variation in the absorption column density persists across all three intensity levels. There is a distinct shift in absorption column density around the orbital phase $\sim 0.4$. As can be seen in the top panel of Fig.~\ref{fig: vela_x1_int_phase_resolved_spectroscopy}, the increase in the absorption column density happens earlier as we move from low to high intensity levels.

\section{Wind modelling}\label{sec: wind_modelling}

\begin{figure}[h]
  \centering
  \includegraphics[width=0.45\textwidth]{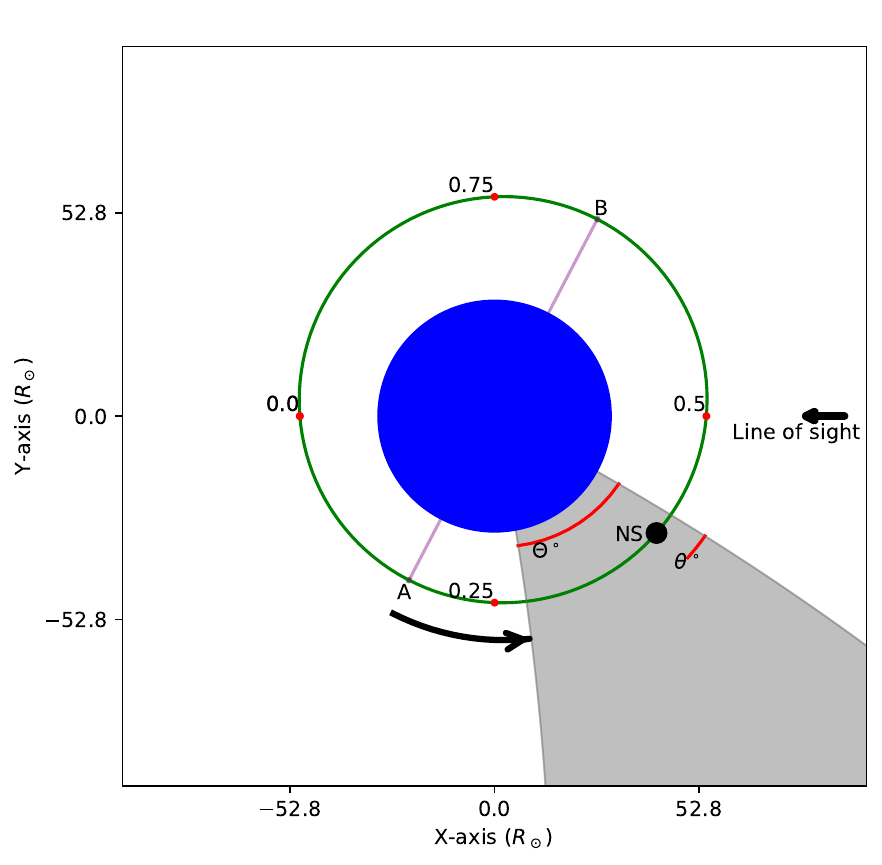}
  \caption{Sketch of~\vela with the photoionisation wake (in grey). NS (in black) is the neutron star at phase 0.4. The companion (in blue) is at the focus of the orbit. The angles $\theta$ and $\Theta$ are the immersion and the opening angles of the neutron star and the wake, respectively. The line AB is the line of apsides, and the line of sight is along the positive X axis.}
  \label{fig: vela_x1_sketch_with_photoionization_wake}
\end{figure}

The asymmetric distribution of the absorption column density after phase 0.4 in both orbital-phase resolved (Fig.~\ref{fig: vela_x1_orbital_phase_resolved_spectroscopy}) and intensity-and-orbital-phase resolved (Fig.~\ref{fig: vela_x1_int_phase_resolved_spectroscopy}) spectroscopy is indicative of the neutron star getting obscured behind dense absorbing matter after that phase. The variation in the overall hardness ratio in Fig.~\ref{fig:vela_x1_total_int_hardness} also suggests the same. In the following sections, we try to understand how the changing geometry of the absorbing medium composed of stellar wind and a trailing photoionisation wake can lead to the observed asymmetric variation in the absorption column density at different intensity levels.

\subsection{Smooth stellar wind with photoionisation wake: A toy model}\label{sec: smooth_wind_and_wake}
The wind velocity profile, $v^\text{SSW}(r)$, of the radiatively driven smooth-stellar wind \citep{smooth_wind_cak75} at a radial distance, $r$, from the centre of the companion star is given as

\begin{equation}
\centering
    v^\text{SSW}(r) = v_\infty  \bigg(1 - \frac{R_*}{r}\bigg)^{\beta^\text{SSW}}
    \label{eq: smooth_wind_velocity}
,\end{equation}

where, $v_\infty$ is the terminal velocity of the stellar wind, $\beta^\text{SSW}$ is the velocity exponent, and $R_*$ is the radius of the companion star. The hydrogen number density for the smooth-stellar wind, $n_H^{\text{SSW}}(r)$, is given as

\begin{equation}
    n_H^{\text{SSW}}(r) =  \frac{\chi_H\dot{M}}{4\pi r^2 v_\text{SSW}(r)}
    \label{eq: ssw_hydrogen_density_profile}
,\end{equation}

where $\chi_H$ is the conversion factor from total mass density to effective hydrogen particle density and $\dot{M}$ is the mass loss rate from the companion star. The absorption column density, ($N_H^{\text{SSW}}$), was obtained by integrating Eq.~\ref{eq: ssw_hydrogen_density_profile} along the line of sight from the position of the neutron star ($r'_{NS}$) in the orbit to the observer as

\begin{equation}
    N_H^{\text{SSW}} = \int_{r'_{NS}}^{\infty} n_H^{\text{SSW}}{(r')} dr'
.\end{equation}

To construct the orbit of Vela X$-$1, we used the orbital ephemeris given in Table~\ref{tab: ephermeris_table}. The inclination of the binary plane is $i > 74^\circ$ (Table~\ref{tab: ephermeris_table}), but for simplicity we assumed the binary plane to be edge-on; that is, $i = 90^\circ$ (similar to \citealt{doroshenko_footprints}). 

Symmetric stellar wind gives rise to a symmetric variation in the absorption column density. Therefore, to account for the asymmetric variation in the absorption column density (top panel of Fig.~\ref{fig: vela_x1_orbital_phase_resolved_spectroscopy} and Fig.~\ref{fig: vela_x1_int_phase_resolved_spectroscopy}), we adopted a simple toy model similar to the one of \citealt{doroshenko_footprints}, consisting of a photoionisation wake formed as a result of intense X-ray emission of the neutron star (Fig.~\ref{fig: vela_x1_sketch_with_photoionization_wake}). In our model, the photoionisation wake is a cone with an opening angle ($\Theta$). The axis of the cone is bent away from the direction of motion of the neutron star by a constant factor called the phase-lag parameter ($\epsilon$) so that it trails the neutron star. Boundary 1 (bd1) and 2 (bd2) of the wake are described by the following co-ordinates:

\begin{equation}
\begin{aligned}
    x_{\text{wake}}^{\text{bd1}}, y_{\text{wake}}^{\text{bd1}} & = r \sin (\epsilon r), r \cos (\epsilon r)
    \\
    x_{\text{wake}}^{\text{bd2}}, y_{\text{wake}}^{\text{bd2}} & = r \sin (\Theta + \epsilon r), r \cos (\Theta + \epsilon r)
\end{aligned}
\label{eq: wake_bd_coordinates}
.\end{equation}

Within the wake, we reduced the velocity of the stellar wind by a constant factor called the velocity reduction factor ($f$) to make the stellar wind denser than the region outside the wake. We also assumed a different velocity exponent ($\beta^\text{wake}$) for the wake. The neutron star sits fully immersed in the wake, making an angle, $\theta$, with respect to the leading edge of the wake, which we call the immersion angle ($\theta$) (Fig.~\ref{fig: vela_x1_sketch_with_photoionization_wake}). Therefore, the velocity, $v^\text{wake}$, and the hydrogen number density, $n^{\text{wake}}_H(r)$, of the wake are given by

\begin{equation}
    v^{\text{wake}}(r) = \frac{v_\infty}{f}  \bigg(1 - \frac{R_*}{r}\bigg)^{\beta^\text{wake}}
\label{eq: wake_velocity}
\end{equation}

and
\begin{equation}
    n^{\text{wake}}_H(r) = \frac{\chi_H \dot{M}}{4\pi r^2 v^\text{wake}(r)}
    \label{eq: wake_column_density}
.\end{equation}

Both Eq.~\ref{eq: wake_bd_coordinates} and Eq.~\ref{eq: wake_column_density} hold for a radial value of $r < r_0$, where $r_0$ is the length scale chosen at a distance where the hydrogen number density inside the wake becomes comparable to the hydrogen number density of smooth stellar wind. 

The equivalent hydrogen number density, $n_H(r)$, is given by

\begin{equation}
    n_H(r) = n^{\text{SSW}}_H(r) + n^{\text{wake}}_H(r)
.\end{equation}

The total absorption column density, ($N_H$), becomes

\begin{equation}
N_H = \int_{r'_{NS}}^{r_{w0}} n^{\text{wake}}_H{(r')} dr' + \int_{r_{w0}}^{\infty} n^{\text{SSW}}_H{(r')} dr'
\label{eq: total_absorption_column_density}
,\end{equation}

where $r_{w0}$ is the extent of the wake along the line of sight.

\subsection{Fitting of observed absorption column density}
\subsubsection{Stellar wind and wake structure for overall Vela X\texorpdfstring{$-$}{-}1 observation}
The four parameters that govern the structure of the wake are the opening angle of the wake ($\Theta$), the phase lag parameter ($\epsilon$), the neutron star immersion angle with respect to the leading edge of the wake ($\theta$), and the velocity reduction factor ($f$). The variable parameters for the wind model are the ratio of the mass loss rate to the terminal velocity ($\dot{M}/v_\infty$), the smooth wind velocity exponent ($\beta^\text{SSW}$), and the wake velocity exponent ($\beta^\text{wake}$). The radius of the companion star is taken as $30R_\odot$~\citep{van_kerkwijk}, while the conversion factor ($\chi_H$) for the absorption column density was calculated as 0.7~\citep{tbabs_wilms_2000}. 

For a given set of four wake parameters, we setup the wake structure and calculate the absorption column density at a given orbital phase value using Eq.~\ref{eq: total_absorption_column_density}. Then we vary the three wind parameters to fit the observed absorption column density. 

\begin{figure}[h]
  \centering
  \includegraphics[width=0.45\textwidth]{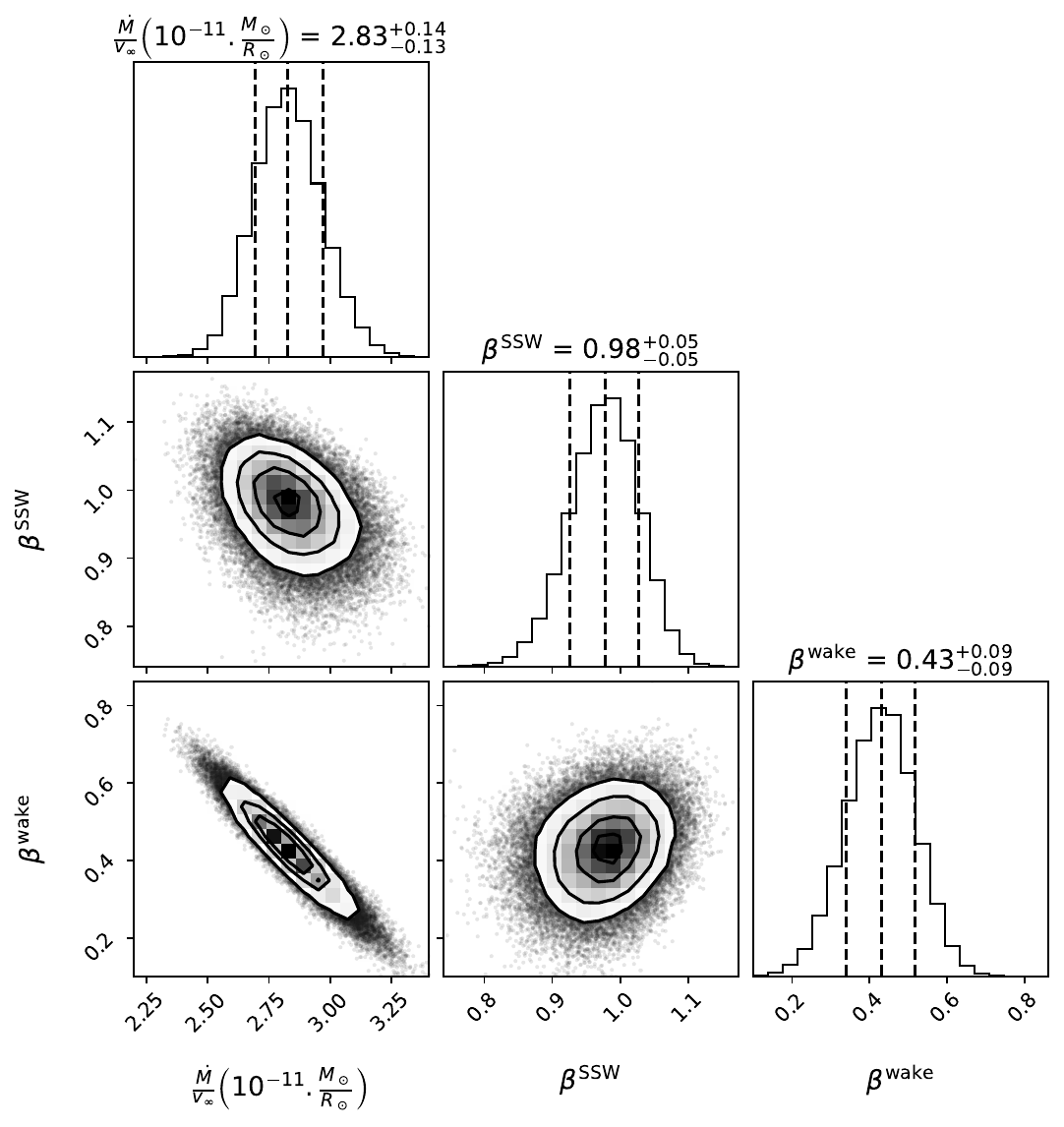}
  \caption{Posterior distribution of the three wind parameters $\dot{M}/v_\infty$, $\beta_{\text{smooth}}$, and $\beta_{\text{wake}}$ obtained from fitting the observed absorption column density to the model predicted absorption column density. The two extreme vertical lines are drawn at a 1$\sigma$ confidence level. The contours correspond to two-dimensional 1$\sigma$, $2\sigma$, and 3$\sigma$ confidence levels.}
  \label{fig: vela_x1_time_avg_wind_para_posterior_distribution}
\end{figure}

\begin{figure}[h]
  \centering
  \includegraphics[width=0.45\textwidth]{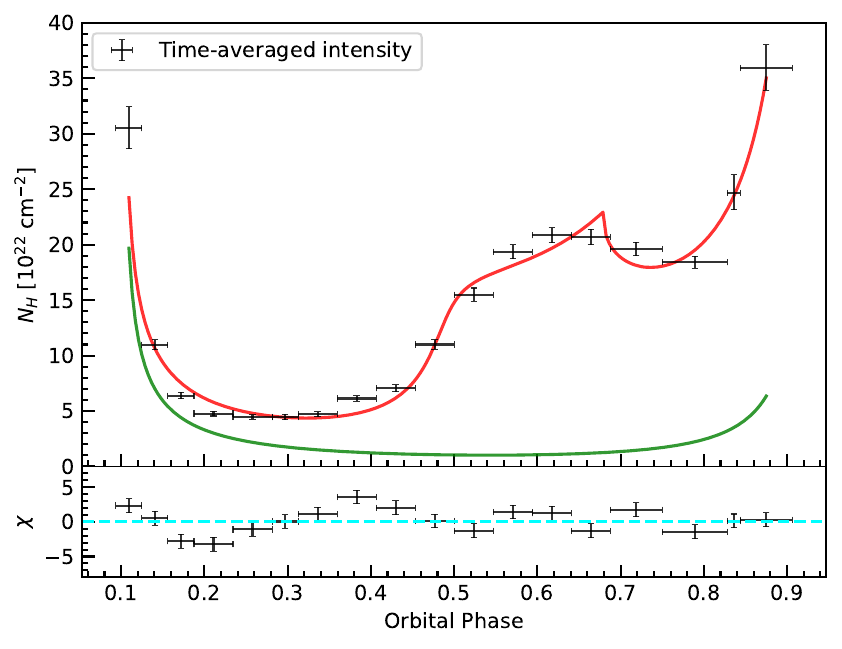}
  \caption{Top panel: Fitting of the orbital-phase resolved absorption column density ($N_H$) to the modified smooth stellar wind model. Black: Observed absorption column density from \maxi/GSC with error bars at a 1$\sigma$ confidence level. Red: Absorption column density with wake. Green: Absorption column density without wake. Bottom panel: Deviation of the data from the best-fitted model.}
  \label{fig:vela_x1_total_int_phase_resolved_nH_vs_phase_fit}
\end{figure}

The wake parameters were varied about their best-fit values to estimate the errors from the chi-square distribution. The errors on the wind parameters were calculated using a Markov chain Monte Carlo sampler (\texttt{Python: emcee}; \citealt{emcee_python}; \citealt{affine_invariance_mcmc}) and their posterior distribution is shown in Fig.~\ref{fig: vela_x1_time_avg_wind_para_posterior_distribution}. The best-fit values for both wind and wake parameters, along with errors estimated at a 1$\sigma$ confidence level, are given in Table~\ref{tab:best_fit_results}. The corresponding best-fit absorption column density predicted by our model is shown in red in Fig.~\ref{fig:vela_x1_total_int_phase_resolved_nH_vs_phase_fit}. The residuals indicate that the model parameters reproduce the data well. From the posterior distribution (Fig.~\ref{fig: vela_x1_time_avg_wind_para_posterior_distribution}), it is evident that there is a strong negative correlation between $\beta^{\text{wake}}$ and $\dot{M}/ v_\infty$. A similar but less strong correlation is observed between $\beta^\text{SSW}$ and  $\dot{M}/ v_\infty$.

\subsubsection{Intensity dependence of wake structure in Vela X\texorpdfstring{$-$}{-}1}
\begin{figure*}
    \begin{subfigure}{0.32\textwidth}
        \centering
        \includegraphics[width=\textwidth]{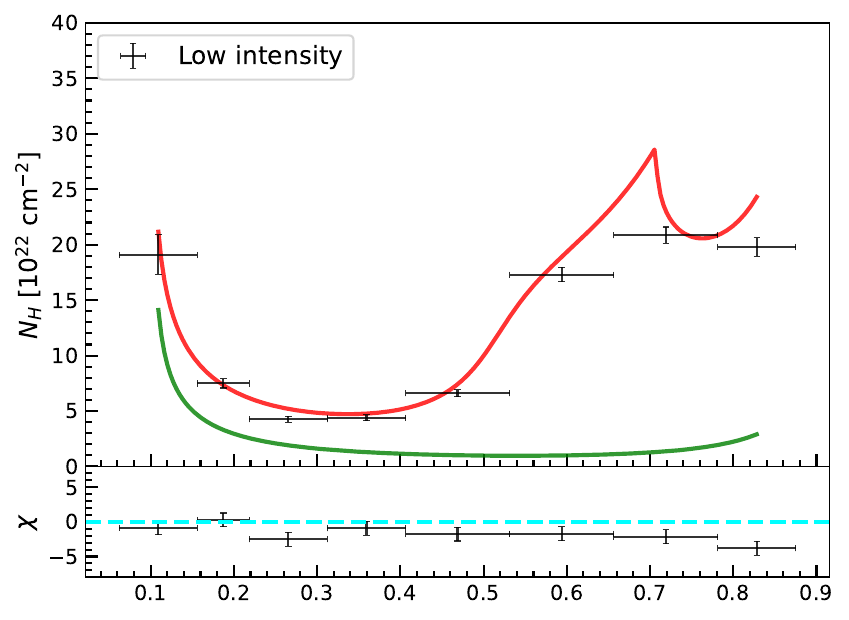}
        \label{fig:sub1}
    \end{subfigure}
    \hfill
    \begin{subfigure}{0.32\textwidth}
        \centering
        \includegraphics[width=\textwidth]{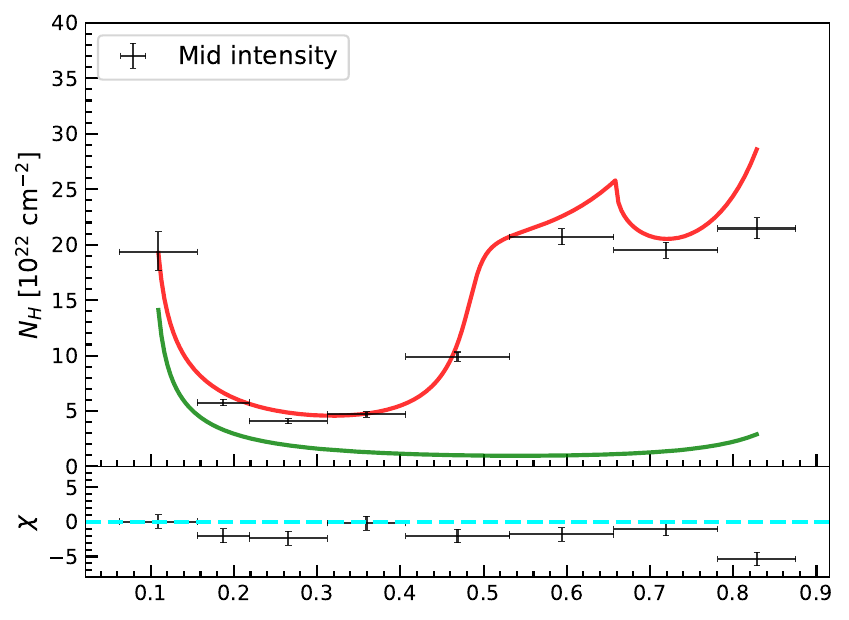}
        \label{fig:sub2}
    \end{subfigure}
    \hfill
    \begin{subfigure}{0.32\textwidth}
        \centering
        \includegraphics[width=\textwidth]{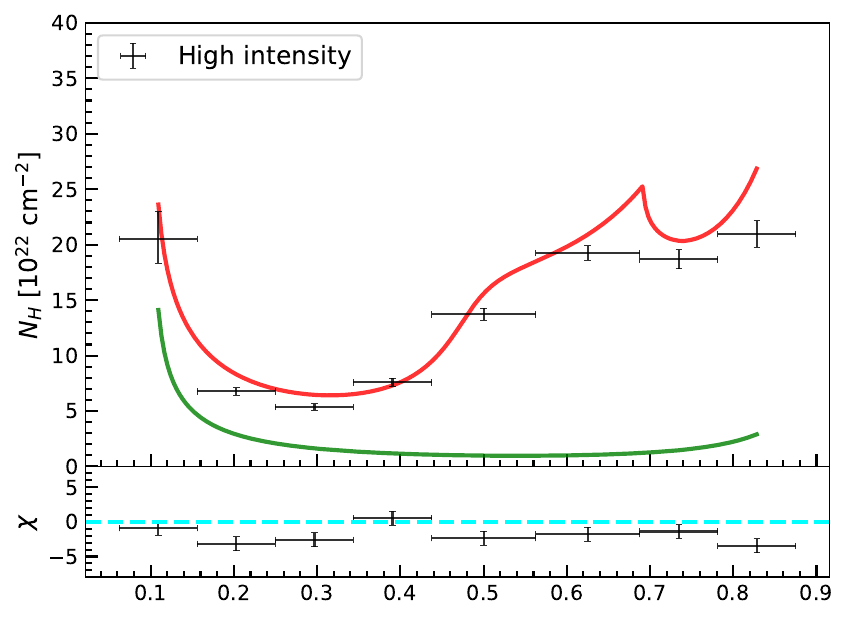}
        \label{fig:sub3}
    \end{subfigure}
    \caption{Top panel: Fitting of the absorption column density derived from the low-, mid-, and high-intensity orbital-phase-resolved spectroscopy to the modified smooth stellar wind model. Black: Observed absorption column density from \maxi/GSC with error bars at 1$\sigma$ confidence level. Red: Absorption column density with wake. Green: Absorption column density without wake. Bottom panel: Deviation of the data from the best-fitted model.
    \label{fig: vela_x1_int012_phase_resolved_fit}}
\end{figure*}

The wind parameters from the companion star are not expected to change over a timescale of decades; therefore, we assumed the same wind parameters for the different intensity levels. We only estimated the wake parameters by freezing the wind parameters to the one obtained from fitting of the observed orbital-phase resolved $N_H$ (Table~\ref{tab:best_fit_results}). The best-fit wake parameters were estimated based on the minimisation of chi-squares for the three intensity levels. The best wake parameters and their errors estimated at a 1$\sigma$ confidence level are given in Table~\ref{tab:best_fit_results}. The best-fit model and the observed absorption column density for the three intensity levels are shown in Fig.~\ref{fig: vela_x1_int012_phase_resolved_fit}. To check the validity of our assumption of freezing the wind parameters to the overall best-fit value, we also varied the wind parameters for the three intensity levels. We found that the resulting best-fit parameters remained within the 1$\sigma$ confidence interval of the value obtained from the overall~\vela data.

\begin{table}
\renewcommand{\arraystretch}{1.5}
\normalsize
\centering
\caption{Time-averaged and intensity-resolved wake and wind parameters of Vela X$-$1 derived from \textrm{MAXI}/GSC data.}
\begin{threeparttable}
\begin{tabularx}{0.5\textwidth}{*{5}{>{\centering\arraybackslash}X}}
\hline
\hline
\multicolumn{1}{l}{\hfill} Parameter        & \multicolumn{1}{c}{Orbital}  & \multicolumn{3}{c}{Intensity \& orbital}                               \\ 
\multicolumn{1}{l}{\hfill}                  & \multicolumn{1}{c}{phase-resolved}     &        \multicolumn{3}{c}{phase-resolved}                                \\ \hline
                                                   &                                        & Low intensity            & Mid intensity             & High intensity                         \\ \cline{3-5} 
$\Theta$\tnote{a}                                  & $51.5^{+5.3}_{-7.4}$                   & $41.0^{+10.5}_{-8.9}$    & $48.0^{+7.7}_{-7.6}$      & $54.0^{+11.7}_{-9.1}$                  \\
$\epsilon$\tnote{b}                                & $7.0^{+4.8}_{-4.0}$                    & $35.0^{+17.7}_{-15.3}$   & $4.02^{+7.9}_{-3.2}$      & $12.5^{+9.6}_{-9.2}$                   \\
$\theta$\tnote{c}                                  & $6.6^{+0.9}_{-1.0}$                    & $6.0^{+1.4}_{-1.3}$      & $5.4^{+1.0}_{-1.1}$       & $10.0^{+2.0}_{-1.8}$                   \\
$f$                                                &$18.5^{+1.9}_{-1.3}$                    & $20.5^{+5.0}_{-4.0}$     & $20.0^{+2.7}_{-2.3}$      & $18.0^{+2.6}_{-2.5}$                   \\
$\dot{M}/v_{\infty}$\tnote{d}                      & $2.83^{+0.14}_{-0.13}$                 & -   & -    & - \\
$\beta^\text{SSW}$                                 & $0.98^{+0.05}_{-0.05}$                 & -   & -    & - \\
$\beta^\text{wake}$                                & $0.43^{+0.09}_{-0.09}$                 & -   & -    & - \\ 
\hline
\hline
\end{tabularx}

Notes: The errors quoted are at a 1$\sigma$ confidence level.

\begin{tablenotes}
    \item[a] Opening angle in degrees.
    \item[b] Phase lag parameter in the units of $ (10^{-4} \, \text{rad} \ R_\odot^{-1})$
    \item[c] Immersion angle in degrees
    \item[d] In the units of $(10^{-11} \, M_\odot \ R_\odot^{-1})$
\end{tablenotes}
\end{threeparttable}
\label{tab:best_fit_results}
\end{table}

\section{Discussion}\label{sec: discussion}

In this work, we have studied the sg-HMXB,~\vela, with almost thirteen years of observations by \maxi/GSC, which has helped us to better investigate the long-term orbital phase dependence of the spectral parameters. The stellar wind from OB-type stars can be characterised by the mass loss rate ($\dot{M}$), the terminal velocity ($v_\infty$), and the velocity exponent ($\beta$) (Eq.~\ref{eq: smooth_wind_velocity}, Eq.~\ref{eq: ssw_hydrogen_density_profile}). In~\vela, the variation in the absorption column density with orbital phase shows a deviation from the symmetric behaviour expected from an isotropic stellar wind (top panels of Fig.~\ref{fig: vela_x1_orbital_phase_resolved_spectroscopy}, Fig.~\ref{fig: vela_x1_int_phase_resolved_spectroscopy}). The asymmetric variation in the absorption column density in~\vela has been observed with different X-ray telescopes~\citep{doroshenko_footprints, review_kretschmar, continuum_cyclotron_absorption_diez_2022, observing_the_onset_diez_2023}. Also, there are hydrodynamic simulations that have been performed to understand the origin of asymmetric variation in the $N_H$ in~\vela \citep{blondin_1990, Manousakis_2012}. A wake-like structure was proposed by \citealt{doroshenko_footprints} to explain the asymmetric behaviour of the absorption column density. The onset of a wake-like structure has been reported by~\citep{continuum_cyclotron_absorption_diez_2022, observing_the_onset_diez_2023} from pointed \textrm{XMM}-Newton and \textrm{NuSTAR} observations. With 13 years of \maxi/GSC data, we confirm the presence and the persistent nature of the wake-like structure (Fig.~\ref{fig: vela_x1_orbital_phase_resolved_spectroscopy}) in \vela, which also exhibits an asymmetric variation in the absorption column density in our line of sight as was previously reported (see Sect. \ref{sec: introduction}). We proposed a wake model in which the geometry of the wake is characterised by an opening angle, immersion angle, phase lag parameter, and a velocity reduction factor (Sect.~\ref{sec: smooth_wind_and_wake}). With a longer dataset, we were able to constrain the wind parameters and also quantitatively calculate the wake parameters, which was not attempted in \citealt{doroshenko_footprints}. The best-fit wake and wind parameters obtained from our analysis are given in Table~\ref{tab:best_fit_results}, where the errors are estimated at a 1$\sigma$ confidence level. The wake parameters are consistent with \citealt{doroshenko_footprints}. The wind parameters agree with previous reports~\citep{Nagase_1986, prinja_1990, Watanabe_2006, falanga_2015}. 

Another characteristic is the intensity fluctuation over the long term exhibited by~\vela in the timescale of the orbital period (Fig.~\ref{fig:vela_x1_long_term_lc_2to20kev}). In other pulsars such as LMC X-4 and Her X-1, the change in intensity on the timescale of a few orbital periods is periodic in nature, varying with a super-orbital period (\citealt{lmc_x4_suporb} for LMC X-4 and \citealt{her_x1_suporb} for Her X-1) that is associated with the precession of binary plane and precession of the accretion disk, respectively. However, no such super-orbital periodicity in~\vela has been reported so far. 

The longer \maxi/GSC data allows us to study the variation in the wake structure with the intensity level. Orbital-phase resolved spectroscopy in the three different intensity states was performed. The main finding of the current work is that the accretion and/or photoionisation wake is present at all intensity levels (Fig.~\ref{fig: vela_x1_int012_phase_resolved_fit}), which is understood from the variation in both the absorption column density with orbital phase (the top panel of Fig.~\ref{fig: vela_x1_int_phase_resolved_spectroscopy}) and the hardness ratios (Fig.~\ref{fig:vela_x1_int_resolved_hardness}). The increase in $N_H$ (Fig.~\ref{fig: vela_x1_int_phase_resolved_spectroscopy}) and the hardness ratio (Fig.~\ref{fig:vela_x1_int_resolved_hardness}) happens earlier at a high intensity, indicating that the geometry of the wake is intensity-dependent. The varying wake parameters for the three intensity levels are given in Table~\ref{tab:best_fit_results} with the errors quoted at a 1$\sigma$ confidence level. The consistent increase in the wake opening angle with the intensity level suggests that the wake structure prefers to be wider for higher intensity levels. Similar values of the velocity reduction factor suggest that the wake remains almost equally dense at all three intensity levels. We do not see such a trend for the phase lag parameter and the immersion angle.

\cite{abalo_2024}, whose findings were published during the review process of this paper, have used the hardness ratio between 2.0 - 4.0 keV and 4.0 - 10.0 keV energy bands to study the X-ray absorbing medium along the line of sight at the individual orbit level. Similar to our analysis, they also show that large-scale structures in Vela X-1, such as the accretion wake trailing the neutron star, significantly influence the observed orbital variation in the absorption patterns, which is consistent with existing literature~\citep{doroshenko_footprints, malacaria_2016}. Their work shows that a considerable number of orbital hardness ratio profiles deviate from the average hardness ratio towards both hard and soft bands, and they attribute this to stochastic effects such as the clumpy nature of the stellar wind and the inherent variability of the source. Complementarily to \cite{abalo_2024}, the current work focuses on the study of the variability of the accretion wake structure with changes in source luminosity. 
  
Currently, because of the edge-on assumption ($i=90^\circ$), we have disregarded the third spatial axis. The immediate modification to the model that one could consider is it being in three dimensions. One could also consider a smooth transition of density at the wake boundaries instead of assuming an abrupt density change, which is beyond the scope of this work.

\begin{acknowledgements}
The authors would like to thank the comments and suggestions from the referee, which helped improve the quality of this manuscript.
This research has made use of \textrm{MAXI} data provided by RIKEN, JAXA and the \textrm{MAXI} team.
\end{acknowledgements}

\bibliographystyle{aa}
\bibliography{references}

\begin{thebibliography}{44}
\expandafter\ifx\csname natexlab\endcsname\relax\def\natexlab#1{#1}\fi

\bibitem[{{Abalo} {et~al.}(2024){Abalo}, {Kretschmar}, {F{\"u}rst}, {Diez}, {El Mellah}, {Grinberg}, {Guainazzi}, {Mart{\'\i}nez-N{\'u}{\~n}ez}, {Manousakis}, {Amato}, {Zhou}, \& {Beijersbergen}}]{abalo_2024}
{Abalo}, L., {Kretschmar}, P., {F{\"u}rst}, F., {et~al.} 2024, \aap, 692, A188

\bibitem[{{Aftab} {et~al.}(2019){Aftab}, {Paul}, \& {Kretschmar}}]{iron_line_paper4}
{Aftab}, N., {Paul}, B., \& {Kretschmar}, P. 2019, \apjs, 243, 29

\bibitem[{{Arnaud}(1996)}]{XSPEC}
{Arnaud}, K.~A. 1996, in Astronomical Society of the Pacific Conference Series, Vol. 101, Astronomical Data Analysis Software and Systems V, ed. G.~H. {Jacoby} \& J.~{Barnes}, 17

\bibitem[{{Becker} {et~al.}(1978){Becker}, {Rothschild}, {Boldt}, {Holt}, {Pravdo}, {Serlemitsos}, \& {Swank}}]{iron_line_paper1}
{Becker}, R.~H., {Rothschild}, R.~E., {Boldt}, E.~A., {et~al.} 1978, \apj, 221, 912

\bibitem[{{Bildsten} {et~al.}(1997){Bildsten}, {Chakrabarty}, {Chiu}, {Finger}, {Koh}, {Nelson}, {Prince}, {Rubin}, {Scott}, {Stollberg}, {Vaughan}, {Wilson}, \& {Wilson}}]{bildsten_1997}
{Bildsten}, L., {Chakrabarty}, D., {Chiu}, J., {et~al.} 1997, \apjs, 113, 367

\bibitem[{{Blondin} {et~al.}(1990){Blondin}, {Kallman}, {Fryxell}, \& {Taam}}]{blondin_1990}
{Blondin}, J.~M., {Kallman}, T.~R., {Fryxell}, B.~A., \& {Taam}, R.~E. 1990, \apj, 356, 591

\bibitem[{{Boerner} {et~al.}(1987){Boerner}, {Hayakawa}, {Nagase}, \& {Anzer}}]{boerner_1987}
{Boerner}, G., {Hayakawa}, S., {Nagase}, F., \& {Anzer}, U. 1987, \aap, 182, 63

\bibitem[{{Castor} {et~al.}(1975){Castor}, {Abbott}, \& {Klein}}]{smooth_wind_cak75}
{Castor}, J.~I., {Abbott}, D.~C., \& {Klein}, R.~I. 1975, \apj, 195, 157

\bibitem[{{Chodil} {et~al.}(1967){Chodil}, {Mark}, {Rodrigues}, {Seward}, \& {Swift}}]{discovery_of_source}
{Chodil}, G., {Mark}, H., {Rodrigues}, R., {Seward}, F.~D., \& {Swift}, C.~D. 1967, \apj, 150, 57

\bibitem[{{Diez} {et~al.}(2023){Diez}, {Grinberg}, {F{\"u}rst}, {El Mellah}, {Zhou}, {Santangelo}, {Mart{\'\i}nez-N{\'u}{\~n}ez}, {Amato}, {Hell}, \& {Kretschmar}}]{observing_the_onset_diez_2023}
{Diez}, C.~M., {Grinberg}, V., {F{\"u}rst}, F., {et~al.} 2023, \aap, 674, A147

\bibitem[{{Diez} {et~al.}(2022){Diez}, {Grinberg}, {F{\"u}rst}, {Sokolova-Lapa}, {Santangelo}, {Wilms}, {Pottschmidt}, {Mart{\'\i}nez-N{\'u}{\~n}ez}, {Malacaria}, \& {Kretschmar}}]{continuum_cyclotron_absorption_diez_2022}
{Diez}, C.~M., {Grinberg}, V., {F{\"u}rst}, F., {et~al.} 2022, \aap, 660, A19

\bibitem[{{Doroshenko} {et~al.}(2013){Doroshenko}, {Santangelo}, {Nakahira}, {Mihara}, {Sugizaki}, {Matsuoka}, {Nakajima}, \& {Makishima}}]{doroshenko_footprints}
{Doroshenko}, V., {Santangelo}, A., {Nakahira}, S., {et~al.} 2013, \aap, 554, A37

\bibitem[{{Falanga} {et~al.}(2015){Falanga}, {Bozzo}, {Lutovinov}, {Bonnet-Bidaud}, {Fetisova}, \& {Puls}}]{falanga_2015}
{Falanga}, M., {Bozzo}, E., {Lutovinov}, A., {et~al.} 2015, \aap, 577, A130

\bibitem[{{Foreman-Mackey} {et~al.}(2013){Foreman-Mackey}, {Conley}, {Meierjurgen Farr}, {Hogg}, {Lang}, {Marshall}, {Price-Whelan}, {Sanders}, \& {Zuntz}}]{emcee_python}
{Foreman-Mackey}, D., {Conley}, A., {Meierjurgen Farr}, W., {et~al.} 2013, {emcee: The MCMC Hammer}, Astrophysics Source Code Library, record ascl:1303.002

\bibitem[{{F{\"u}rst} {et~al.}(2010){F{\"u}rst}, {Kreykenbohm}, {Pottschmidt}, {Wilms}, {Hanke}, {Rothschild}, {Kretschmar}, {Schulz}, {Huenemoerder}, {Klochkov}, \& {Staubert}}]{furst_2010}
{F{\"u}rst}, F., {Kreykenbohm}, I., {Pottschmidt}, K., {et~al.} 2010, \aap, 519, A37

\bibitem[{{Goldstein} {et~al.}(2004){Goldstein}, {Huenemoerder}, \& {Blank}}]{iron_line_paper2}
{Goldstein}, G., {Huenemoerder}, D.~P., \& {Blank}, D. 2004, \aj, 127, 2310

\bibitem[{{Goodman} \& {Weare}(2010)}]{affine_invariance_mcmc}
{Goodman}, J. \& {Weare}, J. 2010, Communications in Applied Mathematics and Computational Science, 5, 65

\bibitem[{{Haberl} \& {White}(1990)}]{Haberl_1990}
{Haberl}, F. \& {White}, N.~E. 1990, \apj, 361, 225

\bibitem[{{HI4PI Collaboration} {et~al.}(2016){HI4PI Collaboration}, {Ben Bekhti}, {Fl{\"o}er}, {Keller}, {Kerp}, {Lenz}, {Winkel}, {Bailin}, {Calabretta}, {Dedes}, {Ford}, {Gibson}, {Haud}, {Janowiecki}, {Kalberla}, {Lockman}, {McClure-Griffiths}, {Murphy}, {Nakanishi}, {Pisano}, \& {Staveley-Smith}}]{nH_calculator_heasarc}
{HI4PI Collaboration}, {Ben Bekhti}, N., {Fl{\"o}er}, L., {et~al.} 2016, \aap, 594, A116

\bibitem[{{Inoue} {et~al.}(1984){Inoue}, {Ogawara}, {Ohashi}, {Waki}, {Hayakawa}, {Kunieda}, {Nagase}, \& {Tsunemi}}]{inoue_1984}
{Inoue}, H., {Ogawara}, Y., {Ohashi}, T., {et~al.} 1984, \pasj, 36, 709

\bibitem[{{Kaper} {et~al.}(1994){Kaper}, {Hammerschlag-Hensberge}, \& {Zuiderwijk}}]{vela_x1_photoionization1994}
{Kaper}, L., {Hammerschlag-Hensberge}, G., \& {Zuiderwijk}, E.~J. 1994, \aap, 289, 846

\bibitem[{{Katz}(1973)}]{her_x1_suporb}
{Katz}, J.~I. 1973, Nature Physical Science, 246, 87

\bibitem[{{Kretschmar} {et~al.}(2021){Kretschmar}, {El Mellah}, {Mart{\'\i}nez-N{\'u}{\~n}ez}, {F{\"u}rst}, {Grinberg}, {Sander}, {van den Eijnden}, {Degenaar}, {Ma{\'\i}z Apell{\'a}niz}, {Jim{\'e}nez Esteban}, {Ramos-Lerate}, \& {Utrilla}}]{review_kretschmar}
{Kretschmar}, P., {El Mellah}, I., {Mart{\'\i}nez-N{\'u}{\~n}ez}, S., {et~al.} 2021, \aap, 652, A95

\bibitem[{{Kreykenbohm} {et~al.}(1999){Kreykenbohm}, {Kretschmar}, {Wilms}, {Staubert}, {Kendziorra}, {Gruber}, {Heindl}, \& {Rothschild}}]{vela_rxte_1999}
{Kreykenbohm}, I., {Kretschmar}, P., {Wilms}, J., {et~al.} 1999, \aap, 341, 141

\bibitem[{{Kreykenbohm} {et~al.}(2008){Kreykenbohm}, {Wilms}, {Kretschmar}, {Torrej{\'o}n}, {Pottschmidt}, {Hanke}, {Santangelo}, {Ferrigno}, \& {Staubert}}]{kreykenbohm_2008}
{Kreykenbohm}, I., {Wilms}, J., {Kretschmar}, P., {et~al.} 2008, \aap, 492, 511

\bibitem[{{Malacaria} {et~al.}(2016){Malacaria}, {Mihara}, {Santangelo}, {Makishima}, {Matsuoka}, {Morii}, \& {Sugizaki}}]{malacaria_2016}
{Malacaria}, C., {Mihara}, T., {Santangelo}, A., {et~al.} 2016, \aap, 588, A100

\bibitem[{{Manousakis} {et~al.}(2012){Manousakis}, {Walter}, \& {Blondin}}]{Manousakis_2012}
{Manousakis}, A., {Walter}, R., \& {Blondin}, J.~M. 2012, \aap, 547, A20

\bibitem[{{Matsuoka} {et~al.}(2009){Matsuoka}, {Kawasaki}, {Ueno}, {Tomida}, {Kohama}, {Suzuki}, {Adachi}, {Ishikawa}, {Mihara}, {Sugizaki}, {Isobe}, {Nakagawa}, {Tsunemi}, {Miyata}, {Kawai}, {Kataoka}, {Morii}, {Yoshida}, {Negoro}, {Nakajima}, {Ueda}, {Chujo}, {Yamaoka}, {Yamazaki}, {Nakahira}, {You}, {Ishiwata}, {Miyoshi}, {Eguchi}, {Hiroi}, {Katayama}, \& {Ebisawa}}]{maxi}
{Matsuoka}, M., {Kawasaki}, K., {Ueno}, S., {et~al.} 2009, \pasj, 61, 999

\bibitem[{{Mihara} {et~al.}(2011){Mihara}, {Nakajima}, {Sugizaki}, {Serino}, {Matsuoka}, {Kohama}, {Kawasaki}, {Tomida}, {Ueno}, {Kawai}, {Kataoka}, {Morii}, {Yoshida}, {Yamaoka}, {Nakahira}, {Negoro}, {Isobe}, {Yamauchi}, \& {Sakurai}}]{maxi_gsc}
{Mihara}, T., {Nakajima}, M., {Sugizaki}, M., {et~al.} 2011, \pasj, 63, S623

\bibitem[{{Nagase} {et~al.}(1986){Nagase}, {Hayakawa}, {Sato}, {Masai}, \& {Inoue}}]{Nagase_1986}
{Nagase}, F., {Hayakawa}, S., {Sato}, N., {Masai}, K., \& {Inoue}, H. 1986, \pasj, 38, 547

\bibitem[{{Paul} \& {Kitamoto}(2002)}]{lmc_x4_suporb}
{Paul}, B. \& {Kitamoto}, S. 2002, Journal of Astrophysics and Astronomy, 23, 33

\bibitem[{{Prinja} {et~al.}(1990){Prinja}, {Barlow}, \& {Howarth}}]{prinja_1990}
{Prinja}, R.~K., {Barlow}, M.~J., \& {Howarth}, I.~D. 1990, \apj, 361, 607

\bibitem[{{Quaintrell} {et~al.}(2003){Quaintrell}, {Norton}, {Ash}, {Roche}, {Willems}, {Bedding}, {Baldry}, \& {Fender}}]{ns_mass}
{Quaintrell}, H., {Norton}, A.~J., {Ash}, T.~D.~C., {et~al.} 2003, \aap, 401, 313

\bibitem[{{Rappaport}(1975)}]{spin_period}
{Rappaport}, S. 1975, \iaucirc, 2869, 2

\bibitem[{{Rikame} {et~al.}(2024){Rikame}, {Paul}, {Sharma}, {Jithesh}, \& {Paul}}]{iron_line_paper5}
{Rikame}, K., {Paul}, B., {Sharma}, R., {Jithesh}, V., \& {Paul}, K.~T. 2024, \mnras, 529, 3360

\bibitem[{{Rodes-Roca} {et~al.}(2015){Rodes-Roca}, {Mihara}, {Nakahira}, {Torrej{\'o}n}, {Gim{\'e}nez-Garc{\'\i}a}, \& {Bernab{\'e}u}}]{rodes_2015}
{Rodes-Roca}, J.~J., {Mihara}, T., {Nakahira}, S., {et~al.} 2015, \aap, 580, A140

\bibitem[{{Sidoli} {et~al.}(2015){Sidoli}, {Paizis}, {F{\"u}rst}, {Torrej{\'o}n}, {Kretschmar}, {Bozzo}, \& {Pottschmidt}}]{sidoli_2015}
{Sidoli}, L., {Paizis}, A., {F{\"u}rst}, F., {et~al.} 2015, \mnras, 447, 1299

\bibitem[{{Sunyaev} \& {Titarchuk}(1980)}]{compst_model}
{Sunyaev}, R.~A. \& {Titarchuk}, L.~G. 1980, \aap, 86, 121

\bibitem[{{Tomida} {et~al.}(2011){Tomida}, {Tsunemi}, {Kimura}, {Kitayama}, {Matsuoka}, {Ueno}, {Kawasaki}, {Katayama}, {Miyaguchi}, {Maeda}, {Daikyuji}, \& {Isobe}}]{maxi_ssc}
{Tomida}, H., {Tsunemi}, H., {Kimura}, M., {et~al.} 2011, \pasj, 63, 397

\bibitem[{{van Kerkwijk} {et~al.}(1995{\natexlab{a}}){van Kerkwijk}, {van Paradijs}, \& {Zuiderwijk}}]{orbital_period}
{van Kerkwijk}, M.~H., {van Paradijs}, J., \& {Zuiderwijk}, E.~J. 1995{\natexlab{a}}, \aap, 303, 497

\bibitem[{{van Kerkwijk} {et~al.}(1995{\natexlab{b}}){van Kerkwijk}, {van Paradijs}, {Zuiderwijk}, {Hammerschlag-Hensberge}, {Kaper}, \& {Sterken}}]{van_kerkwijk}
{van Kerkwijk}, M.~H., {van Paradijs}, J., {Zuiderwijk}, E.~J., {et~al.} 1995{\natexlab{b}}, \aap, 303, 483

\bibitem[{{Verner} {et~al.}(1996){Verner}, {Ferland}, {Korista}, \& {Yakovlev}}]{Vern_cs}
{Verner}, D.~A., {Ferland}, G.~J., {Korista}, K.~T., \& {Yakovlev}, D.~G. 1996, \apj, 465, 487

\bibitem[{{Watanabe} {et~al.}(2006){Watanabe}, {Sako}, {Ishida}, {Ishisaki}, {Kahn}, {Kohmura}, {Nagase}, {Paerels}, \& {Takahashi}}]{Watanabe_2006}
{Watanabe}, S., {Sako}, M., {Ishida}, M., {et~al.} 2006, \apj, 651, 421

\bibitem[{{Wilms} {et~al.}(2000){Wilms}, {Allen}, \& {McCray}}]{tbabs_wilms_2000}
{Wilms}, J., {Allen}, A., \& {McCray}, R. 2000, \apj, 542, 914

\end{thebibliography}

\end{document}